\documentclass[11pt,a4paper]{article}
\usepackage{jheppub}
\usepackage{graphicx}
\usepackage{epsfig}
\title{\Large\bf Discrete torsion, de Sitter tunneling vacua and AdS brane:\\ 
${\mathbf{U(1)}}$ gauge theory on ${\mathbf{D_4}}$-brane and an effective curvature}
\author{Abhishek K. Singh,}
\author{K. Priyabrat Pandey,}
\author{Sunita Singh and }
\author{Supriya Kar}
\affiliation  {\large Department of Physics \& Astrophysics\\ \large University of Delhi, New Delhi 110 007, India}
\emailAdd{aksingh@physics.du.ac.in}
\emailAdd{kppandey@physics.du.ac.in}
\emailAdd{sunita@physics.du.ac.in}
\emailAdd{skkar@physics.du.ac.in}

\abstract{The $U(1)$ gauge dynamics on a $D_4$-brane is revisited, with a two form, to construct an effective curvature theory in a second order formalism. We exploit the local degrees in a two form, and modify its dynamics in a gauge invariant way, to incorporate a non-perturbative metric fluctuation in an effective $D_4$-brane. Interestingly, the near horizon $D_4$-brane is shown to describe an asymptotic Anti de Sitter (AdS) in a semi-classical regime. Using Weyl scaling(s), we obtain the emergent rotating geometries leading to primordial de Sitter (dS) and AdS vacua in a quantum regime. Under a discrete transformation, we re-arrange the mixed dS patches to describe a Schwazschild-like dS (SdS) and a topological-like dS (TdS) black holes. We analyze SdS vacuum for Hawking radiations to arrive at Nariai geometry, where a discrete torsion forms a condensate. We perform thermal analysis to identify Nariai vacuum with a TdS. Investigation reveals an AdS patch within a thermal dS brane, which may provide a clue to unfold dS/CFT. In addition, the role of dark energy, sourced by a discrete torsion, in the dS vacua is investigated using Painleve geometries. It is argued that a D-instanton pair is created by a discrete torsion, with a Big Bang/Crunch, at the past horizon in a pure dS. Nucleation, of brane/anti-brane pair(s), is qualitatively analyzed to construct an effective space-time on a $D_4$-brane and its anti brane. Analysis re-assures the significant role played by a non-zero mode, of NS-NS two form, to generalize the notion of branes within a brane.}

\arxivnumber{1303.4344}

\begin{document}
\maketitle
\flushbottom


\def\ra{{\rightarrow}}
\def\a{{\alpha}}
\def\b{{\beta}}
\def\l{{\lambda}}
\def\eps{{\epsilon}}
\def\T{{\Theta}}
\def\t{{\theta}}
\def\co{{\cal O}}
\def\car{{\cal R}}
\def\caf{{\cal F}}
\def\cs{{\Theta_S}}
\def\pr{{\partial}}
\def\tri{{\triangle}}
\def\na{{\nabla }}
\def\S{{\Sigma}}
\def\s{{\sigma}}
\def\sp{\vspace{.15in}}
\def\hs{\hspace{.25in}}

\newcommand{\be}{\begin{equation}} \newcommand{\ee}{\end{equation}}
\newcommand{\bea}{\begin{eqnarray}}\newcommand{\eea}
{\end{eqnarray}}

\section{Introduction}
The holographic approach to explore quantum gravity is possibly a noble attempt to generalize existing string models of gravity. In particular, the success underlying AdS and conformal field theory (CFT) duality \cite{maldacena,witten} may provide a powerful tool to revisit the gauge theories on a $D$-brane and its near horizon black hole geometries. Along its progress, a variant of AdS/CFT correspondence, underlying a dS space brane-world between two AdS geometries, has been passionately observed \cite{hawking-maldacena-strominger}. Importantly, a holographic duality has been conjectured \cite{strominger-ds,bousso-maloney-strominger,klemm} between a quantum gravity on a $dS_p$ to an euclidean CFT on $S^{(p-1)}$. The dS/CFT correspondence has been supported by the correlators of a massive scalar field. However, the dual CFT in general may be non-unitary. Nevertheless, there has been revival of interest to explore dS geometries \cite{bousso-hawking}-\cite{banks-fiol-morisse} underlying certain aspects of quantum gravity.

\sp
\noindent
On the other hand, the present limit on the vacuum energy density leading to a small positive cosmological constant motivates an intense research to revisit dS with renewed perspectives. In the context, an effective $D$-brane formulation, underlying a type II superstring, may be helpful to unfold some aspects of dS geometries and their tunnelings. Generically, a dS black hole is bounded by a cosmological horizon, which makes it very different than an AdS or an asymptotically flat black hole. dS defines a maximally symmetric space. Kerr dS black hole is known to be the most general known gravity described by the vacuum Einstein field equations in presence of a positive cosmological constant. 

\sp
\noindent
Importantly, a quantum dS space has been investigated \cite{parikh-wilczek}-\cite{louis} to unfold some of its mysteries in various contexts. It has been conjectured that an asymptotic dS is bounded by a dS entropy. However, an asymptotic dS does not possess a spatial infinity unlike to that in an asymptotic AdS. A globally time-like killing vector is not available to an observer in SdS black hole which is defined with a negative gravity mass. Nevertheless, a TdS black hole possesses a positive gravity mass and an observer is on a spatial phase beyond a cosmological horizon. Apparently, some of the conceptual issues in SdS may be avoided in a TdS vacuum, which may serve an intermittent quantum phase underlying a tunneling between the meta-stable SdS and a stable AdS. As a part of this paper, we attempt to explore a plausible tunneling scenario underlying an effective $D_4$-brane dynamics and geometries. However, the topic needs further attention in the present literatures.

\sp
\noindent
In the recent past, there have been attempts to construct various near horizon $D$-brane effective geometries including dS.
A zero mode of a NS-NS two form, in a gauge invariant combination with an electromagnetic field, has been shown to describe a nonlinear $U(1)$ gauge symmetry  \cite{seiberg-witten}. Interestingly, the AdS/CFT duality has been explored in presence of a non-zero mode of a NS-NS two form in bulk AdS 
and its zero mode in a $U(1)$ gauge theory on a $D$-brane \cite{li}. It has been conjectured that a non-zero mode, in principle, would like to modify the invariant gauge curvature, which would like to enforce a non-constant, non-commutative, parameter on a $D$-brane. However, a non-zero mode has not been explored to its strength to construct an effective near horizon geometriy on a $D$-brane. Nevertheless, a zero mode has been exploited to illustrate the significance of non-linear electro-magnetic charges leading to various near horizon deformations on a $D$-brane \cite{gibbons}-\cite{kahle-minasian}. The emerging geometries on a $D$-brane, alternately underlie a non-commutative space-time on its world-volume. Furthermore, a generic non-commutative space-time has been investigated to address a notion of emergent gravity from the $U(1)$ gauge fields \cite{yang}-\cite{cai}.

\sp
\noindent
In the context, a symplectic two form in the NS-NS sector of type II superstring theories possibly urge an attention in the recent time. A two form is sourced by a string charge density vector and is known to play a significant role to describe a D-brane dynamics and its underlying geometries. It is believed to enlighten some of the geometric transitions \cite{kar-ds,kar} underlying D-brane deformation geometries and their thermodynamics. The deformations on a D-brane may reveal insights into the meta-stable dS vacua and their tunneling phenomena. In fact, a two form generates a torsion in the $U(1)$ gauge theory on a $D$-brane. The gauge theoretic torsion may be modified appropriately to define a geometric torsion in a second order formalism. 

\sp
\noindent
The geometric torsion may formally be identified with the torsion in the Einstein-Cartan Theory (ECT). Interestingly, torsion geometries in string backgrounds leading to spinning branes have been in investigated \cite{vasilic-vojinovic,vasilic,maier,maier-falciano}. It has been argued that a two form is related to a torsion itself rather than to its potential. In fact, a discrete torsion in a $D$-brane has been addressed by using a projective representation of the orbifold group \cite{douglas}. Presumably, a discrete torsion may be viewed as a choice of orbifold group action on a two form. Subsequently, a discrete torsion has been incorporated into the gauge theory on $D$-branes and has carefully been analyzed in various contexts \cite{subir-koushik}-\cite{douglas-fiol}. 

\sp
\noindent
In the paper, we attempt to formulate an emergent fourth order effective curvature on a $D_4$-brane by exploiting the $U(1)$ gauge dynamics of a two form on its world-volume. In particular, we construct an appropriate covariant derivative to modify a $H_3$ to a geometric ${\cal H}_3$, using a irreducible tensor connection sourced by a two form in the frame-work \cite{kpss1,kpss2}. The $U(1)$ gauge invariance under a two form transformation is shown to incorporate non-perturbative metric correction to a $D_4$-brane. The effective curvature theory describes a propagating torsion in the near horizon brane geometry underlying a second order formalism. Subsequently, we investigate some of the emergent quantum vacua underlying a dS, a TdS and an AdS, in addition to a semi-classical emergent gravity on an effective $D_4$-brane. As a result, we show that a near horizon $D_4$-brane in a semi-classical regime is identified with an asymptotic $AdS_2\times S^3$.

\sp
\noindent
In the quantum regime, a torsion is shown to play a significant role to describe various emergent geometries. It is argued that a discrete torison may provide a clue to the origin of dark energy in an emergent dS and AdS. The quantum patches are analyzed in Painleve coordinates to qualitatively describe the origin of a brane-Universe with a Big Bang at the past horizon of a pure dS. Nucleation, of brane/anti-brane pair, is analyzed to construct an effective curvature underlying a near horizon $D_4$-brane. Interestingly, the non-zero modes of two form seem to generalize the notion of branes within a brane \cite{douglas-cargese} in the frame-work. Most importantly the tunneling, of dS vacuum to a stable AdS via a number of meta-stable vacua underlying an effective $D_4$-brane, reveals an AdS patch within a thermal dS brane. 

\sp
\noindent
We plan to organize the paper as follows. An effective fourth order curvature, underlying a two form gauge theory on a $D_4$-brane, is constructed in section 2. The effective theory is shown to describe a propagating torsion in an effective $D_4$-brane. In section 3, we focus on a near horizon $D_4$-brane geometry in the large r limit. The primary objective of this paper may seen to explore an emergent dS and an AdS quantum geometries in section 4. There, we discuss a pair creation of brane/anti-brane Universes by a discrete torsion at the Big Bang. In section 5, we conclude with a summary and out-line a few plausible research problems in the frame-work.

\section{Effective curvature formulation on a ${\mathbf{D}}$-brane}

\subsection{${\mathbf{U(1)}}$ Gauge theory on a ${\mathbf{D_4}}$-brane}
The $U(1)$ gauge theory on a $D_4$-brane is primarily described by a gauge field $A_{\mu}$. In presence of a constant background metric $g_{\mu\nu}$ on a $D_4$-brane, the linear gauge dynamics is given by
\be
S_{\rm A}= -{1\over{4C_1^2}}\int d^5x\ {\sqrt{-g}}\ F_{\alpha\beta}F^{\alpha\beta}\ ,\label{gauge-1}
\ee 
where $C_1^2=(4\pi^2g_s){\alpha'}^{1/2}$ denotes the gauge coupling. In a Coulomb gauge ($A_0=0$ and $\nabla\cdot{\mathbf A}=0$), the theory  describes an electromagnetic field $F_{\alpha\beta}=(\nabla_{\alpha}A_{\beta}-\nabla_{\beta}A_{\alpha})$. We fix the metric signature to ($+,-,-,-,-$) in the one form theory. The field strength is invariant under the gauge transformations 
\be
A_{\alpha} \rightarrow A_{\alpha} + \delta A_{\alpha}\ ,\qquad {\rm where}\quad \ \delta A_{\alpha}=  \partial_{\alpha}\epsilon\ .\label{gauge-2}
\ee
Alternately, the $U(1)$ gauge theory on a $D_4$-brane may be revisited by a two form $B_{\mu\nu}$ alone, whose field strength is Poincare dual to that of the electromagnetic field. Poincare duality can interchange the metric signature between the original and the dual. Thus, the metric signature is naturally fixed to ($-,+,+,+,+$) in the two form gauge theory. The two form gauge dynamics is given by
\bea
S_{\rm B}=- {1\over{12C_2^2}}\int d^5x\ {\sqrt{-g}}\ H_{\mu\nu\lambda}H^{\mu\nu\lambda}\ ,\qquad {\rm where}\quad H_{\mu\nu\lambda}=3\nabla_{[\mu}B_{\nu\lambda ]}\ ,\label{gauge-3}
\eea
where $C_2^2=(8\pi^3g_s){\alpha'}^{3/2}$ defines an appropriate gauge coupling in a two form theory. The gauge conditions may be worked out to yield
\be
B^{0i}=0\qquad {\rm and}\quad \nabla_iB^{ij}=0\ .\nonumber\\
\ee
It is straightforward to check that the dual three form field $H_{\mu\nu\lambda}$ remains invariant under the $U(1)$ gauge transformations
\bea
B_{\mu\nu} \rightarrow B_{\mu\nu} + \delta B_{\mu\nu}\ ,\qquad {\rm where}\quad {1\over{\sqrt{2\pi\alpha'}}}\ \delta B_{\mu\nu}=  \partial_{\mu}\epsilon_{\nu}-\partial_{\nu}\epsilon_{\mu}\ .\label{gauge-4}
\eea
We recall that the $A_{\mu}$ gauge theory is linear, and is governed by three local degrees, on a $D_4$-brane. However, a linear gauge theory qualitatively differs from its non-linear cousin in their global properties. Both linear and nonlinear gauge theories differ in their gauge transformation, though their equivalent role on a $D$-brane has been established by Seiberg and Witten \cite{seiberg-witten}. In principle, the gauge equivalence between the respective Poincare dual theories, underlying a two form, may be argued on a $D_4$-brane. The realization is indeed supported by a fact that the dual of $F_2$  and ${\cal F}_2^{nz}=(F_2+ (2\pi\alpha')^{-1}B^z_2)$ on a $D_4$-brane is described by a non-linear gauge theoretic torsion $H_3$. In the paper, we attempt to construct a (geometric) ${\cal H}_3$ by exploiting the local degrees in a two form to describe a near horizon $D_4$-brane. As a result, we obtain two effective curvature theories, respectively, underlying a (propagating) gauge theoretic torsion and a geometric torsion in the frame-work. The geometric torsion dynamics is further explored to revisit the Big Bang and the Big Crunch at the beginning of the Universe, respectively on a brane and its anti-brane.

\subsection{Modified {$\mathbf{{\cal H}_3}$}: Geometric torsion}
In this section, we explore an effective curvature theory on a $D_4$-brane by exploiting its two-form gauge connections under its inherent $U(1)$ gauge symmetry. Non-linearity being a global notion, the local degrees remain unchanged. We construct an appropriate covariant derivative 
${\cal D}_{\mu}$ on the brane using the two form connections. It modifies a three form field strength and incorporates an effective curvature underlying a weak gravity perturbation into the brane gauge dynamics. A priori, the operation of the covariant derivative on a two form may be defined as
\bea
{\cal D}_{\lambda}B_{\mu\nu}&=&\partial_{\lambda}B_{\mu\nu} - 
\left (\Gamma^{\rho}_{\lambda\mu} + {{\mathbf\Gamma}_{\lambda\mu}}^{\rho}\right ) B_{\rho\nu} + \left ( \Gamma^{\rho}_{\lambda\nu} + {{\mathbf\Gamma}_{\lambda\nu}}^{\rho}
\right )B_{\rho\mu}\nonumber\\
&=&\nabla_{\lambda}B_{\mu\nu} - {{\mathbf\Gamma}_{\lambda\mu}}^{\rho}B_{\rho\nu} + {{\mathbf\Gamma}_{\lambda\nu}}^{\rho}B_{\rho\mu}\ ,\label{gauge-5}
\eea
\vspace{-.4in}
\bea
{\rm where}\qquad {{\mathbf\Gamma}_{\mu\nu}}^{\rho}&=&g^{\rho\lambda}{\mathbf\Gamma}_{\mu\nu\lambda}\qquad\qquad\qquad\qquad\qquad\qquad\qquad\qquad \quad\qquad\qquad\qquad{}\nonumber\\
&=&-{1\over2} g^{\rho\lambda} \partial_{[\lambda} B_{\mu\nu ]}\ .\label{gauge-5a}
\eea
The irreducible tensor connections, sourced by a symplectic two form, is identified with the three form gauge curvature ${{\mathbf\Gamma}_{\mu\nu}}^{\rho}=-{1\over{2}}{H_{\mu\nu}}^{\rho}$. We modify $H_3$ to ${\cal H}_3$ by incorporating the two-form gauge connections.
Naively, the modified three form may be constructed in a perturbation gauge theory using a covariant derivative. Analysis reveals
an iterative incorporation of $B_2$-corrections, to all orders, in the covariant derivative otherwise defined in a gauge theory. As a result, 
the exact covariant derivative in a perturbative gauge theory may seen to define a non-perturbative covariant derivative in a second order formalism. Then a non-perturbative covariant derivative defined in a geometric realization is given by
\be
{\cal D}_{\lambda}B_{\mu\nu}=\nabla_{\lambda}B_{\mu\nu} + {1\over2}{{{\cal H}}_{\lambda\mu}}^{\rho}B_{\rho\nu} - {1\over2}{{\cal H}_{\lambda\nu}}^{\rho}B_{\rho\mu}\ .\label{gauge-5b}
\ee
All order $B_2$-corrections in a gauge theory yields
${{\mathbf\Gamma}_{\mu\nu}}^{\rho} \rightarrow {{\tilde{\mathbf\Gamma}}_{\mu\nu}}^{\rho}=-{1\over{2}}{{\cal H}_{\mu\nu}}^{\rho}$, which is identified with a geometric torsion in a second order formalism. Explicitly, we write
\bea
{\cal H}_{\mu\nu\lambda}&=&3{\cal D}_{[\mu}B_{\nu\lambda ]}\nonumber\\
&=&3\nabla_{[\mu}B_{\nu\lambda ]} + 3{{\cal H}_{[\mu\nu}}^{\alpha}
{B^{\beta}}_{\lambda ]}\ g_{\alpha\beta}
\nonumber\\ 
&=&H_{\mu\nu\lambda} + \left ( H_{\mu\nu\alpha}{B^{\alpha}}_{\lambda} + \rm{cyclic\; in\;} \mu,\nu,\lambda \right )\ +\ H_{\mu\nu\beta} {B^{\beta}}_{\alpha} {B^{\alpha}}_{\lambda} + \dots \;\ .\label{gauge-6}
\eea
In fact the two form dynamics in a first order formalism may be viewed as a torsion dynamics on an effective $D_p$-brane in a second order formalism. In other words, a gauge torsion $H_3$ in a perturbation theory (\ref{gauge-3}) may alternately be described by a fourth order curvature tensor $K_{\mu\nu\lambda\rho}$. Interestingly, a geometric torsion ${\cal H}_3$ defined appropriately by a non-perturbative covariant derivative (\ref{gauge-5b}) may equivalently be described by an analogous curvature tensor ${\cal K}_{\mu\nu\lambda\rho}$ in a geometric theory. We postpone the detailed discussion to the section 2.4.

\sp
\noindent
On the other hand, a modified ${\cal H}_3$ is not gauge invariant under a two form gauge transformations (\ref{gauge-4}), though $H_3$ is a gauge invariant. Importantly, the $U(1)$ gauge invariance in a modified action defined with a ${\cal H}_3$ enforces a nontrivial metric fluctuation 
in the theory. The non-perturbative fluctuation, underlying a $U(1)$ gauge invariance, turns out to be governed by the fluxes and is given by
\bea
f_{\mu\nu}^{nz}&=&C\ {\cal H}_{\mu\alpha\beta}\ {{\bar{\cal H}}^{\alpha\beta}{}}_{\nu}\nonumber\\
&\approx&C\ H_{\mu\alpha\beta}\ {{{\bar{\cal H}}}^{\alpha\beta}{}}_{\nu}\ ,\label{gauge-7}
\eea
where $C$ is an arbitrary constant and ${\bar{\cal H}}_{\mu\nu\lambda}= (2\pi\alpha'){\cal H}_{\mu\nu\lambda}$. Hence the five dimensional brane-world dynamics, underlying a $U(1)$ gauge dynamics, may equivalently be described either by a $H_3$ in a gauge theory or by a ${\cal H}_3$ in a geometric theory. Interestingly both the torsions, $H_3$ and ${\cal H}_3$, may equivalently be described by the effective curvatures $K_{\mu\nu\lambda\rho}$ and ${\cal K}_{\mu\nu\lambda\rho}$, respectively in a second order formalism. Contrary to a constant background metric underlying a gauge theoretic torsion dynamics, a geometric torsion is associated with a background metric fluctuation. In principle, the geometric formulation may seen to incorporate a mass term $m^2B_{\mu\nu}B^{\mu\nu}$ in the gauge theory. Thus a massive two form in five dimensions becomes sensible, when it is viewed from a ten dimensional superstring theory or an eleven dimensional M-theory. However, within the realm of a gauge theory on a $D_4$-brane, a $B_{\mu\nu}$ field is mass-less.

\sp
\noindent
In the context, a geometric torsion may appropriately be compared with a torsion the Einstein Cartan Theory (ECT). However, a torsion in the frame-work is completely anti-symmetric which differs from that in ECT. In particular, a torsion in ECT is not completely antisymmetric in its indices. It defines a con-torsion tensor which in turn modifies a covariant derivative in ECT. Interestingly the connections 
${{\cal H}_{\mu\nu}}^{\lambda}$ may naively be generalized to define a con-torsion tensor. Then the con-torsion tensor takes a form: 
\be
{{\tilde{\mathbf\Gamma}}_{\mu\nu}}^{\lambda} = {1\over2}\left ({{\cal H}_{\mu\nu}}^{\lambda} -{{{\cal H}_{\nu}}^{\lambda}}_{\mu} + {{\cal H}^{\lambda}}_{\mu\nu}\right )\ .\label{gauge-8}
\ee 
In fact a torsion in ECT may be identified with a traceless part of the con-torsion. Under a consistent truncation of the local degree in scalar field theory, the torsion in ECT may correspond to a geometric torsion in the frame-work. Then the con-torsion tensor reduces to a torsion in absence of a scalar field in ECT and the boundary term vanishes there. 

\subsection{Geometric ${\mathbf{{\tilde{\cal F}}_2}}$}
We revisit a non-linear ${\cal F}_2$, in a $U(1)$ gauge theory in presence of a dynamical two form on a D-brane. A non-linear ${\cal F}_2$ in an effective description incorporates a non-zero mode of NS-NS two form in the open string theory. It is known that a zero mode of NS-NS two form in the open string theory can couple appropriately to a boundary gauge field to yield a gauge theoretic non-linear ${\cal F}^z_2$ on a $D$-brane
\cite{seiberg-witten}. A non-zero mode stays in the string bulk along with the metric tensor. Together, they govern an effective space-time dynamics in target space. Nevertheless, a $D$-brane away from its world-volume along its transverse directions can purview various near horizon geometries leading to extremal black holes and black branes \cite{gibbons,mars,ishibashi}. In fact these non-trivial geometries are likely to be governed by an underlying ten dimensional type II superstrings or an eleven dimensional M-theory. Thus a non-zero mode becomes significant to govern an effective geometry on a $D$-brane.

\sp
\noindent
A priori, we recall the Poincare duality, if any, between a geometric ${\cal H}_3$ and a gauge theoretic ${\cal F}_2$, underlying the $U(1)$ gauge symmetry on a $D_4$-brane. We qualitatively address a plausible generalization of the non-linear ${\cal F}^z_{\alpha\beta}$. Naively, the dual of ${\cal H}_3$ may be worked out to yield
\be
{{\tilde{\cal F}}}^{\alpha\beta}\rightarrow \left (\ {\cal F}_z^{\alpha\beta}- {1\over{\sqrt{2\pi\alpha'}}}\ {{\epsilon^{\alpha\beta\mu\nu\lambda}}\over{3\sqrt{-g}}}\  {{H}_{[\mu\nu}}^{\sigma} {B^{\delta}}_{\lambda ]}\ g_{\sigma\delta}\ \right )\ .\label{gauge-9}
\ee
A ${\cal H}_3$ alone is not a $U(1)$ gauge invariant. However, its combination as lorentz scalar is indeed a gauge invariant. As a result, a ${{\tilde{\cal F}}}_2$ in eq.(\ref{gauge-9}) seems to break the $U(1)$ gauge invariance. The problem lies in the fact that the ${\cal F}_2$ is field theoretic and its dual ${\cal H}_3$ is geometric. They indeed require correct formalism. Thus, an apparent problem is resolved when we work with the dual of ${\cal H}_3$ using an appropriate covariant derivative (\ref{gauge-5b}) in a second order formalism. Then, a geometric two form field strength may appropriately be given by
\bea
{\tilde {\cal F}}_{\alpha\beta}&=&{\cal D}_{\alpha}A_{\beta} -{\cal D}_{\beta}A_{\alpha}\nonumber\\
&=&\left ( {\cal F}^z_{\alpha\beta} + {{\cal H}_{\alpha\beta}}^{\delta}A_{\delta}\right )\ .\label{gauge-10}
\eea
The $U(1)$ gauge invariance in the action defined with a lorentz scalar ${\tilde{\cal F}}^2$ may also seen to incorporate a metric fluctuation in the formalism. The emerging geometric fluctuation (\ref{gauge-7}) in its dual description may be given by
\be
f_{\mu\nu}^{nz}={\tilde C}\ (2\pi\alpha')^2\ {\tilde{\cal F}}_{\mu\alpha}{{\tilde{\cal F}}^{\alpha}{}}_{\nu}\ ,\label{gauge-11}
\ee
where ${\tilde C}$ is an arbitrary constant. The geometric field strengths (\ref{gauge-9}) and (\ref{gauge-10}) differ from each other in their respective forms coupling to a gauge torsion and a geometric torsion. A two form field strength naturally reduces to a gauge two form in a second order formalism when ${\cal H}_3\rightarrow H_3$. In other words, the dual of $H_3$ may also be obtained directly by use of an appropriate covariant derivative (\ref{gauge-5}). The gauge theoretic two form field strength becomes 
\bea
{\cal F}_{\alpha\beta}&=&{\cal D}_{\alpha}A_{\beta} -{\cal D}_{\beta}A_{\alpha}\nonumber\\
&=&\left ( F^z_{\alpha\beta} + {H_{\alpha\beta}}^{\delta}A_{\delta}\right )\ .\label{gauge-12}
\eea
It is evident that a gauge torsion modifies the electromagnetic field, and its underlying $U(1)$ gauge symmetry, to their non-linear cousin. The $U(1)$ gauge invariance of the electro-magnetic field is restored for its non-linear field strength for a topological coupling of the gauge field to a gauge torsion. The conserved charges in the non-linear gauge theory differ from that of linear theory by its global mode. For a topological torsion, the modified ${\cal F}_{\alpha\beta}$ may be identified with a gauge invariant ${\cal F}^z_{\alpha\beta}$. Generically, ${\cal F}_2$ may be viewed as a generalization of a non-linear field strength ${\cal F}^z_2$.

\sp
\noindent
A priori, the presence of two form, coupled to a gauge torsion, may appear to break the $U(1)$ gauge invariance of the electromagnetic theory. Nevertheless, the gauge invariance in presence of ${\cal H}_3$ in the geometric theory re-assures the invariance in its dual non-linear ${\cal F}_2$ gauge theory. In other words, the $U(1)$ gauge symmetry is elevated to its non-linear, which in turn is preserved in the theory. The non-linear gauge invariance may also seen to be restored for an adiabatic electromagnetic field. In the case, the non-linear two form field strength may be approximated to yield:
\be
{\cal F}_{\alpha\beta}=\left ( {\cal F}^z_{\alpha\beta} - {1\over2} {{H}_{\alpha\beta}}^{\delta}F_{\delta\rho} X^{\rho}\right )\ .\label{gauge-13}
\ee
In fact the electromagnetic field in presence of a gauge torsion, coupled either to a two form or to an one form, yields a generic non-linear ${\cal F}_2$. It is unlike to that in ${\cal H}_3$ expression (\ref{gauge-6}), which is solely described by a dynamical two form. In fact, the one form gauge theory is not independent of its dual theory due to the underlying non-trivial effective metric correction on the brane-world in the formalism. As a result, a two form gauge theory may be a good description due to its independent nature than the usual one form dynamics on a $D_4$-brane. Thus a geometric torsion dynamics primarily leads to an effective curvature description on a $D_4$-brane which underlie a non-linear $U(1)$ gauge theory.

\subsection{Torsion dynamics}
In the section, we focus on a subtle aspect of an effective space-time theory underlying an irreducible curvature in a second order formalism. In fact the notion of effective curvature is dictated in a first order formalism underlying a field theoretic torsion connections. The $U(1)$ gauge invariance, in ${\cal H}_3$ dynamics, further enforces a geometric notion in the frame-work. A generic curvature, underlying an effective space-time, may be constructed at the expense of a non-linear gauge curvature on a $D_4$-brane. The commutator of the gauge covariant exact derivative on a scalar field $\phi(x)$ yields:
\be
\Big [ {\cal D}_{\mu}\ ,\ {\cal D}_{\nu} \Big ]\phi(x) = {{\cal H}_{\mu\nu}}^{\lambda} \partial_{\lambda} \phi(x)\ .\label{gauge-14}
\ee
It hints at the presence of a geometric torsion propagating on the brane-world in an effective description underlying a $U(1)$ gauge theory. The non-vanishing commutator acting on a scalar field in a gauge theory may a priori lead to an apparent paradox when compared with that in Einstein's gravity. A scalar field, being linear, can not incorporate any curvature on a Riemannian manifold. However, a dynamical scalar field in a field theoretic description is dual to a higher form gauge field, which can incorporate non-linear curvatures. For instance, the significance of a geometric torsion on a $D_p$-brane may be identified with $p\ge2$. The torsion turns out to be topological on a $D_2$-brane in a second order formalism. However, the scalar field (dual to an one form) may seen to be dynamical in the gauge theory. The local torsion becomes vital in a gauge theoretic description on a $D_3$-brane and its higher dimensional branes underlying type II superstring theories.

\sp
\noindent
In the context, we compute a commutator between the generalized gauge covariant derivatives, $D_{\mu}\equiv ({\cal D}_{\mu} -A_{\mu})$, defined in a classical theory to explore the possibility of a dynamical one form in the frame-work. The commutator acts on an one form to yield:
\be
\Big [ D_{\mu}\ ,\ D_{\nu} \Big ]A_{\lambda}=\ {\cal F}_{\nu\mu}A_{\lambda}\ +\ {{\cal H}_{\mu\nu}}^{\rho}\ {\cal D}_{\rho}A_{\lambda}
\ +\ {{\cal K}_{\mu\nu\lambda}}^{\rho}A_{\rho} \ +\ {{\cal L}_{\mu\nu\lambda}}^{\rho}A_{\rho} \ .\label{gauge-15}
\ee 
\vspace{-.4in}
\bea
{\rm Where}\qquad {4{\cal K}_{\mu\nu\lambda}}^{\rho}&=&2\partial_{\mu}{{\cal H}_{\nu\lambda}}^{\rho} -2\partial_{\nu} {{\cal H}_{\mu\lambda}}^{\rho} 
+ {{\cal H}_{\mu\lambda}}^{\sigma}{{\cal H}_{\nu\sigma}}^{\rho}-{{\cal H}_{\nu\lambda}}^{\sigma}{{\cal H}_{\mu\sigma}}^{\rho}\;
\;\qquad\qquad\qquad\qquad\qquad\nonumber\\
{\rm and}\qquad 2{{\cal L}_{\mu\nu\lambda}}^{\rho}&=&\left ( \Gamma^{\rho}_{\mu\sigma} {{\cal H}^{\sigma}}_{\nu\lambda} +\Gamma^{\sigma}_{\nu\lambda} {{\cal H}^{\rho}}_{\mu\sigma} - \Gamma^{\sigma}_{\mu\lambda} {{\cal H}^{\rho}}_{\nu\sigma} -\Gamma^{\rho}_{\nu\sigma} {{\cal H}^{\sigma}}_{\mu\lambda}\right )\ .\label{gauge-16}
\eea
For Minkowskian space-time, the fourth order curvature tensor ${{\cal L}_{\mu\nu\lambda}}^{\rho}$ becomes trivial. The commutator hints at a non-trivial effective curvature ${\cal K}_{\mu\nu\lambda\rho}$ underlying a geometric torsion ${\cal H}_3$ dynamics in the frame-work. In other words, the brane-world dynamics may be approximated by an effective curvature theory governed by a dynamical torsion ${\cal H}_3$. The fourth order reducible tensor ${\cal K}_{\mu\nu\lambda\rho}$ is antisymmetric under an exchange of indices within a pair, $i.e.$ under $\mu\leftrightarrow\nu$ and $\lambda\leftrightarrow\rho$. However, it is not symmetric under an exchange of its first pair of indices with the second, 
$i.e.\ (\mu\nu)\leftrightarrow (\lambda\rho)$. Hence, it differs from the Riemannian tensor $R_{\mu\nu\lambda\rho}$ and may be identified with a generalized curvature tensor. Nevertheless, for a topological torsion ${\cal K}_{\mu\nu\lambda\rho}\rightarrow R_{\mu\nu\lambda\rho}$. Other relevant curvature tensors are worked out to yield:
\bea
&&4{\cal K}_{\mu\nu}= -\left (2\partial_{\lambda}{{\cal H}^{\lambda}}_{\mu\nu} +
{{\cal H}_{\mu\rho}}^{\lambda}{{\cal H}_{\lambda\nu}}^{\rho}\right )
\nonumber\\
{\rm and} &&{\cal K}= -{1\over{4}}{\cal H}_{\mu\nu\lambda}{\cal H}^{\mu\nu\lambda}
\ .\label{gauge-17}
\eea
Within a gauge choice leading to a topological torsion, the second order curvature tensor formally identifies with a metric fluctuation ${\cal K}_{\mu\nu}= - {1\over{4C}}f_{\mu\nu}^{nz}$. It re-assures a non-perturbative nature of $f_{\mu\nu}^{nz}$, induced by a topological torsion, on a $D$-brane.

\subsection{Effective metric in presence of ${\mathbf{B_2^{nz}}}$}
Since closed strings are tangential to a $D$-brane, Einstein's gravity decouples from its world-volume. Then an effective space-time geometry solely supported by a non-linear field strength sustains on a $D$-brane. An effective $D$-brane near horizon geometry, for a constant NS-NS two form $B^z_2$, has already been established through an open string metric \cite{seiberg-witten}. In a non-linear $U(1)$ gauge theory on a $D$-brane, the effective metric may be given by
\be
G^z_{\mu\nu} = \left ( g_{\mu\nu} - {\bar{\cal F}}^z_{\mu\lambda}g^{\lambda\rho}{\bar{\cal F}}^z_{\rho\nu}\right )\ ,\label{gauge-12a}
\ee
where $g_{\mu\nu}$ is a constant non-degenerate metric. In absence of cosmological constant ${\tilde\Lambda}$, the $G^z_{\mu\nu}\rightarrow g_{\mu\nu}$, which may be identified with an asymptotic vacuum. In a pure gauge, the open string metric may be expressed as
\be
G^z_{\mu\nu} = \left ( g_{\mu\nu} - B^z_{\mu\lambda}g^{\lambda\rho} B^z_{\rho\nu}\right )\ .\label{gauge-18}
\ee
A zero mode can not be gauged away by any local field transformation, which in turn is known to play a vital role to incorporate a non-linear description on a $D$-brane in absence of Einstein's gravity. Unfortunately, a non-zero mode of NS-NS two form remains in string bulk and hence dissociates from a $D$-brane. However, with a dual gauge theoretic description on a $D_4$-brane, it may be feasible to see the effect of a non-zero mode via an effective metric in the frame-work. We recall a significant role played by a zero mode in the open string metric. The emergent metric may take an identical form to that of an open string metric. In presence of a dynamical two form, leading to a geometric torsion, the emergent metric in a perturbative gravity description may arguably be given by
\bea
{{\tilde G}}_{\mu\nu}&=&\left ( G^z_{\mu\nu} - B_{\mu\lambda}{B^{\lambda}{}}_{\nu}\right )\nonumber\\
&=&\left ( g_{\mu\nu} - B_{\mu\lambda}{B^{\lambda}{}}_{\nu}\right )\ .\label{gauge-19}
\eea
Needless to mention that a $F_2$ in the open string metric can be gauged away and hence a $B_2$ includes both zero as well as non-zero modes in theory. Eq.(\ref{gauge-19}) re-assures the fact that $G^z_{\mu\nu}$ and $g_{\mu\nu}$ are two equivalent vacua on a $D$-brane, respectively, described by a non-linear and linear one form gauge theories. Contrary to the open string metric, ${\tilde G}_{\mu\nu}$ is not unique in an effective space-time due to a two form gauge transformation. However, one may construct a very large number of distinct effective geometries on a $D_4$-brane, which presumably correspond to vacua underlying a landscape scenario in string theory discussed with fluxes \cite{kklt,quevedo,friedmann-stanley,louis}. In the context, the ansatz corresponding to a two form  may be argued to be associated with some dynamical constants. In certain limit(s), the near horizon $D_4$-brane may seen to describe an interesting class of rotating black holes and branes underlying dS and AdS. Nevertheless, we restrict to gauge invariant curvatures in this paper.

\sp
\noindent
Using a dual field source on a $D_4$-brane, an effective open string metric (\ref{gauge-18}) may be worked out to yield:
\be
G^z_{\mu\nu}\rightarrow G^{nz}_{\mu\nu}=\left ( g_{\mu\nu} - \ {\bar H}_{\mu\lambda\rho} 
{H^{\lambda\rho}{}}_{\nu}\right )\ .\label{gauge-20}
\ee 
It is evident that an emergent metric generalizes to include a gauge invariant curvature, due to a dynamical two form in a gauge theory, on a $D_4$-brane. However, the established role of a zero mode appears to be missing on a $D$-brane. Thus, the effective metric (\ref{gauge-20}) may appropriately be modified to include a zero mode. In a field theoretic description, an effective metric on a generic $D$-brane may be re-expressed as:
\bea
G_{\mu\nu}&=&\left ( g_{\mu\nu} - {\bar{\cal F}}^z_{\mu\lambda}{{\bar{\cal F}}^{z\lambda}{}}_{\nu} - H_{\mu\lambda\rho}
{{\bar H}^{\lambda\rho}{}}_{\nu}\right )\nonumber\\
&=&\left ( G^z_{\mu\nu} - H_{\mu\lambda\rho}{{\bar H}^{\lambda\rho}{}}_{\nu}\right )\ .\label{gauge-21} 
\eea
On the other hand, a geometric torsion is intrinsically associated with a nontrivial metric $f_{\mu\nu}^{nz}$. Thus, a field theoretic and a geometric descriptions are defined, respectively, with a constant metric and a propagating metric. Alternately, they may be viewed as two equivalent descriptions underlying a gauge theory on a $D$-brane. In other words, a metric fluctuation emerges in the disguise of an invariant curvature sourced by a propagating geometric torsion on a $D_4$-brane. The geometric torsion on a $D_4$-brane is an artifact in gauge theory due to the redundancy in its gauge degree of freedom. In a gauge fixed theory the torsion disappears. Nevertheless, the gauge degrees make it feasible to analyze a non-perturbative correction to a flat brane-world in the frame-work. Though, we focus on a two form effective dynamics in the formalism, a global mode of NS-NS two form may also be present on a generic $D_4$-brane. A global mode incorporates a shift in the vacuum configuration $g_{\mu\nu}\rightarrow G^z_{\mu\nu}$. In other words, a  global mode of NS-NS two form signifies a non-zero cosmological constant. Incorporating an appropriate metric correction, the effective metric in a geometric formalism on a $D_4$-brane may alternately be described either by a two form field strength or by a three form field strength. It is given by 
\bea
{\tilde G}_{\mu\nu}^{D_4}&=&G^z_{\mu\nu}\ +\ f_{\mu\nu}^{nz}\nonumber\\ 
&=&\left ( G^z_{\mu\nu}\ +\ C\ {\bar{\cal H}}_{\mu\lambda\rho}\ {{\cal H}^{\lambda\rho}{}}_{\nu}\right )\nonumber\\
&=&\left ( G^z_{\mu\nu}\ +\ {\tilde C}\ {\bar{\cal F}}_{\mu\lambda}{{\bar{\cal F}}^{\lambda}{}}_{\nu}\right )
\ .\label{gauge-22}
\eea 
The gauge invariant curvatures make an effective metric unique. A global mode plays a significant role in the geometric description, through its coupling to an electromagnetic field $F_2$. However, the local modes enter through a geometric torsion into an effective metric and hence they incorporate the notion of an dynamical gravity on an effective $D_4$-brane. 

\subsection{${\mathbf{D_p}}$-brane effective action}
The dynamics on an effective $D_4$-brane may be approximated by an irreducible curvature obtained in a classical prescription. In a second order formalism, the geometro-dynamics of a torsion ${\cal H}_3$ on an effective $D_4$-brane may be approximated by
\be
S_{\rm D_4}^{\rm eff}= {1\over{3C_4^2}}\int d^5x {\sqrt{-{\tilde G}}}\ \left ( {\cal K}- {\tilde\Lambda}\right )\ ,\label{gauge-23}
\ee
where $C_4^2=(8\pi^3g_s){\alpha'}^{3/2}$ is a constant and ${\tilde G}=\det {\tilde G}_{\mu\nu}$. The cosmological constant ${\tilde\Lambda}$, in the geometric action, is sourced by a zero mode in the theory. One may generalize the effective space-time curvature theory, underlying a $U(1)$ gauge symmetry of a dynamical two form, to an arbitrary $D_p$-brane. Thus for an effective $D_p$-brane with $p\ge 2$, a curvature theory may be given by
\be
S_{\rm D_p}^{\rm eff}={1\over{3C^2_p}}\int d^{p+1}x {\sqrt{-{\tilde G}}}\ \left ( {\cal K}^{(p+1)} - {\tilde\Lambda}\right )\ ,\label{gauge-24}
\ee
where $C^2_p$ is a constant. The appropriate covariant derivative (\ref{gauge-5b}) on an effective metric in the formalism yields
\vspace{-.2in}
\be
{\cal D}_{\lambda}{\tilde G}_{\mu\nu}={\cal D}_{\lambda}G^z_{\mu\nu} + {\cal D}_{\lambda}f^{nz}_{\mu\nu}=0\ .\label{gauge-24a}
\ee
The variation of the action under a two form is worked out to yield the non-linear equations of motion and they are given by
\bea
&&\Big [\left ( \partial_{\sigma}g^{\rho\nu} {\cal H}^{\sigma\mu\lambda}- \partial_{\sigma}g^{\rho\mu}{\cal H}^{\sigma\nu\lambda} + \partial_{\sigma}g^{\rho\sigma}\ {\cal H}^{\mu\nu\lambda}\right )\qquad\qquad\qquad\qquad\qquad {}\nonumber\\
&&\qquad-{1\over2} g_{\alpha\beta}\partial_{\sigma}g^{\alpha\beta}\left ( g^{\rho\nu}{\cal H}^{\sigma\mu\lambda} - 
g^{\rho\mu}{\cal H}^{\sigma\nu\lambda} + g^{\rho\sigma}{\cal H}^{\mu\nu\lambda}\right )\Big ] B_{\lambda\rho}\nonumber\\
&&\qquad\qquad +g^{\mu\rho}\partial_{\lambda}\left ( B_{\sigma\rho}{\cal H}^{\lambda\sigma\nu}\right ) 
- g^{\nu\rho}\partial_{\lambda}\left (B_{\sigma\rho}{\cal H}^{\lambda\sigma\mu}
\right ) +g^{\rho\lambda}\partial_{\lambda}\left (B_{\sigma\rho}{\cal H}^{\mu\nu\sigma}\right )\nonumber\\
&&\qquad\qquad\qquad +g^{\mu\rho}\left (\partial_{\lambda}
B_{\sigma\rho}\right ){\cal H}^{\lambda\sigma\nu} - g^{\nu\rho}\left (\partial_{\lambda}B_{\sigma\rho}\right )
{\cal H}^{\lambda\sigma\mu}\nonumber\\
&&\qquad\qquad\qquad\qquad +{1\over2}g^{\mu\lambda}\left (\partial_{\lambda}B_{\sigma\rho}\right ){\cal H}^{\sigma\rho\nu}-
{1\over2}g^{\nu\lambda}\left (\partial_{\lambda}B_{\sigma\rho}\right )
{\cal H}^{\sigma\rho\mu}= 0\ .\label{gauge-25}
\eea
Interestingly, a torsion connection takes an identical form to that of Christoeffel connection in Riemannian geometry. Note that a 
${\cal H}_3$, iteratively, incorporates all higher orders in $B_2$ field. The non-perturbative notion of a geometric torsion in the frame-work is thought provoking and needs further attention. 

\sp
\noindent
Alternately in a gauge theoretic description, underlying a second order formalism, the relevant two form equations are worked out to yield
\be
\partial_{\lambda}{\mathbf \Gamma}^{\lambda\mu\nu} + {1\over2} \Big ( g^{\alpha\beta}\partial_{\lambda}\ g_{\alpha\beta}\Big )
{\mathbf \Gamma} ^{\lambda\mu\nu}=0\ . \label{gauge-26}
\ee
An ansatz for a two form is used to construct a ${\cal H}_3$ propagating in a ${{\cal K}_{\mu\nu\lambda}}^{\rho}$ curvature theory. The second order formalism leading to a fourth order dynamical curvature tensor may also be worked out with a field theoretic torsion when ${\cal H}_3\rightarrow H_3$. As a result, ${\cal K}\rightarrow K$ and a metric fluctuation vanishes, which are in conformity with a gauge theoretic description on a $D$-brane. Then an effective space-time curvature theory reduces to an effective gauge curvature theory and may well be derived from the geometric action (\ref{gauge-23}). An effective action, underlying a gauge torsion, in a second order formalism may be approximated by
\bea
S_{\rm D_p}^{\rm eff}&=&{1\over{3C_p^2}}\int d^{p+1}x {\sqrt{-g}}\ K^{(p+1)}\nonumber\\
&\rightarrow&{1\over{3C_p^2}}\int d^{p+1}x {\sqrt{-G^z}}\ \left ( K^{(p+1)}-\Lambda\right )\ .\label{gauge-27}
\eea
A trivial $\Lambda$ in the gauge theoretic description is re-assured by the absence of a zero mode in its vacuum configuration. It may as well be seen due to the missing coupling of a global mode to a gauge torsion (\ref{gauge-5a}). Thus a zero mode shifts the vacuum energy density to a non-zero value and hence a generic $D_p$-brane effective action may be identified with the second expression in eq.(\ref{gauge-27}). 

\sp
\noindent
To enhance, an understanding on the effective curvature formulation, let us focus on a $D_2$-brane, where a torsion turns out to be topological.
However, for a topological torsion, the generic fourth order curvature reduces to the Riemmanian tensor. From the perspective of a geometric torsion (\ref{gauge-6}), a topological torsion may be constructed when a gauge theoretic torsion $H_3$ cancels its local degrees against its dynamical coupling to a NS-NS two form. Thus a geometric theory (\ref{gauge-24}) on a $D_2$-brane can accommodate the local degrees of two form which are otherwise freezed in a gauge theory (\ref{gauge-27}). It may be interesting to construct a BTZ black hole \cite{btz} in an effective geometric curvature theory on a $D_2$-brane. Thus a curvature theory, for a non-propagating geometric torsion, may reduce to Einstein-Hilbert action in three dimensions. On the other hand, a two form freezes it local degree completely on a $D_2$-brane and hence the gauge theory may not describe a BTZ black hole.

\sp
\noindent
The energy-momentum tensor in a $D_p$-brane gauge theory (\ref{gauge-27}) may be computed using the $U(1)$ gauge invariant curvatures underlying a second order formalism. It is worked out to yield:
\be
T_{\mu\nu}={1\over{6}}\left ({\Lambda-K}\right )G^z_{\mu\nu} - {1\over4} H_{\mu\rho\lambda} {H^{\rho\lambda}}_{\nu}\ .\label{gauge-28}
\ee
In a gauge choice $\Lambda=(3/{\pi\alpha'})+K$, the energy momentum tensor sources an effective metric (\ref{gauge-21}) in a gauge theory. The effective metric is indeed a generalization of the open string metric, in presence of a gauge theoretic torsion, for an effective $D$-brane. A non-zero energy-momentum tensor $T_{\mu\nu}$, underlying a two form $U(1)$ gauge theory, may re-assure the dynamical nature of the cosmological constant in the formalism. The presence of a non-zero positive constant between $\Lambda$ and $K$ in the gauge, ensures that the vacuum energy density underlying the effective geometries on a $D$-brane can take all three (positive, zero and negative) small values. Thus, an effective geometry as viewed by a $D$-brane can as well corresponds to a dS or to an AdS in addition to its intrinsically flat geometry. Furthermore, the trace of $T_{\mu\nu}$ in a gauge theory, $i.e.\ T=\Lambda+(p-5)/(2\pi\alpha')$ ensures a new phenomenon. It is evident that the energy-momentum tensor is modified in presence of a generic vacuum energy density. Thus, a cosmological constant breaks the classical conformal invariance otherwise present in an underlying gauge theory on a $D$-brane. Interestingly, the conformal symmetry is restored when $\Lambda= (5-p)/(2\pi\alpha')$ on a $D_p$-brane.

\sp
\noindent
A geometric torsion in the effective action (\ref{gauge-23}) may seen to be sourced by an appropriate energy-momentum tensor. It is worked out to yield:
\bea
{\tilde T}_{\mu\nu}&=&{1\over{6}}\left ({{\tilde\Lambda}-{\cal K}}\right ){\tilde G}_{\mu\nu} - {1\over{8C\pi\alpha'}}f_{\mu\nu}^{nz}\nonumber\\
&=&{1\over6}\left ( {{\tilde\Lambda}-{\cal K}}\right )G^z_{\mu\nu} + \left ( {{\tilde\Lambda-{\cal K}}\over{6}} - {1\over{8C\pi\alpha'}}\right )f_{\mu\nu}^{nz}\ . \label{gauge-29}
\eea
The trace of energy-momentum tensor becomes 
\be
{\tilde T}={1\over6}\left ({5-p}\right ){\cal K} + {1\over6}\left ({p+1}\right )\Lambda\ .\label{gauge-30}
\ee
With a gauge ${\tilde\Lambda}=(3/{\pi\alpha'})+{\cal K}$, which is identical to that used in $H_3$ gauge theory, the energy-momentum tensor becomes
\bea
(2\pi\alpha'){\tilde T}_{\mu\nu}&=&\left ( G^z_{\mu\nu}\ +\ \left ( C-{1\over4}\right )\ {\bar{\cal H}}_{\mu\lambda\rho} {{\cal H}^{\lambda\rho}}_{\nu}\right )\nonumber\\
&=&\left ( {\tilde G}_{\mu\nu}\ - {1\over4} {\bar{\cal H}}_{\mu\lambda\rho} {{\cal H}^{\lambda\rho}}_{\nu}\right )\ .\label{gauge-31}
\eea
The trace of energy-momentum, ${\tilde T}={\tilde\Lambda}+(p-5)/(2\pi\alpha')$, in a geometric formulation satisfies an identical relation to that in a $H_3$ gauge theory. A priori, the ${\tilde T}_{\mu\nu}$ may seen to incorporate an additional metric correction to an effective metric (\ref{gauge-22}). However, a geometric torsion in an effective metric may be redefined appropriately to imply that ${\tilde T}_{\mu\nu}$ is indeed a precise source to the effective metric ${\tilde G}_{\mu\nu}$. It re-assures an underlying non-perturbative nature of an effective metric in the frame-work.
For instance, an emergent metric sourced by a ${\tilde T}_{\mu\nu}$ with $C=-(1/4)$ on a $D_p$-brane for arbitrary $p$ within a domain of type II superstring theories is given by 
\bea
{\tilde G}_{\mu\nu}^{D_p}&=&\left ( G^z_{\mu\nu}\ -{1\over2}\ {{\bar{\cal H}}_{\mu\lambda\rho}}\ {{\cal H}^{\lambda\rho}{}}_{\nu}\right )\nonumber\\
&=&\left ( G^z_{\mu\nu}\ -{1\over2}\ {{\bar{\cal F}}_{\mu\lambda}}\ {{\cal F}^{\lambda}{}}_{\nu}\right )
\ .\label{gauge-32}
\eea 
Most importantly, the generalized geometries (\ref{gauge-22}) and (\ref{gauge-32}) incorporate the non-zero modes in addition to the zero modes of a NS-NS two form. A torsion may seen to incorporate conserved quantities, $i.e.$ angular momentum and/or a charge, on a brane-world. Interestingly, the torsion brane-world geometries become significant in the Planck energy scale and may be identified with a primordial rotating and/or charged black holes and their higher dimensional generalizations.

\subsection{Two form ansatz}
A priori, an effective $D_4$-brane may seen to be influenced by a generic torsion in the formalism. Nevertheless, we argue that the torsion completely decouples to yield a stable brane both in semi-classical and quantum regimes. The two form ansatz may be expressed as:
\bea
B_{t\psi}&=& B_{r\psi}={{b}\over{(2\pi\alpha')^{1/2}}}\ ,\qquad\qquad\qquad {}\nonumber\\
\qquad B_{\theta\psi}&=&\ {{{\tilde P}^3}\over{(2\pi\alpha')^{3/2}}}\ (\sin^2 \psi\ \cot\theta)\nonumber\\
{\rm and}\qquad B_{\psi\phi}&=&{{P^3}\over{(2\pi\alpha')^{3/2}}}\ (\sin^2\psi\ \cos\theta)\ ,\label{ads-1}
\eea
where $(b,P,{\tilde P})>0$ may signify the (dynamical) constants in the formalism. They may be identified with the conserved quantities defined in an asymptotic regime underlying an effective brane geometry. The range of the angular coordinates are: $0<\psi<\pi\;\ ,\quad 0<\theta<\pi\;\; {\rm and}\quad 0\le\phi< 2\pi$. They describe an $S^3$-vaccum configuration, in absence of a cosmological constant, leading to an $S^3$-line element:
\be
d\Omega^2_3= \left ( d\psi^2\ +\  \sin^2\psi\ \left [ d\theta^2\ +\ \sin^2 \theta\ d\phi^2\right ] \right )\ .\label{ads-1a}
\ee
A geometric torsion, sourced by a two form in a gauge theory, is worked out to yield
\bea
{{\cal H}_{\theta\phi}}^{\psi}&=&(2\pi\alpha')^{-1}\ {{P^3}\over{r^2}}\ (\sin^2\psi \sin \theta)\nonumber\\
{\rm and}\;\; {{\cal H}_{\theta\phi}}^t&=&-{{\cal H}_{\theta\phi}}^r\nonumber\\ 
&=&(2\pi\alpha')^{-3/2}\ {{bP^3}\over{r^2}} (\sin^2\psi \sin \theta)\ .\label{ads-2}
\eea
Note that a geometric torsion, in the gauge choice, is independent of ${\tilde P}^3$ while the two form depends on it. Thus, a gauge invariant emergent metric (\ref{gauge-32}) shall be independent of the conserved charge ${\tilde P}^3$, which in turn may be identified with a topological charge. In principle, ${\tilde P}^3$ can modify another charge, in association, for its global properties in a gauge theory. In addition, the generated ansatz for the geometric torsion may seen to disappear in the limit $r\rightarrow\infty$, where the conserved quantities $(b,P)$ are measured. Thus in a gauge fixed two form theory, there shall be no local torsion which in turn is in agreement with the superstring theory and its underlying Einstein's gravity.

\section{Semi-classical regime: Near horizon ${\mathbf{D_4}}$-brane}

\subsection{${\mathbf{(D{\bar D})_4}}$ system: Nucleation of $D_5$-brane}
In this section, we explore a near horizon $D_4$-brane geometry in the large $r$ limit underlying a $U(1)$ gauge invariant effective curvature scalar (\ref{gauge-23}). The mirror symmetry $r\rightarrow -r$ in the near horizon $D_4$-brane is analyzed in presence of its angular momentum to hint at the existence of a ${\bar D}_4$-brane in the frame-work. The geometric torsion may seen to yield an effective anti-brane notion leading to a nucleation of a $D_5$-brane presumably in an eleven dimensional M-theory. In particular, we begin with an $U(1)$ gauge theory governed by a two form on a $D_4$-brane. We consider the one form gauge fields, if any from higher dimensions, to be in a pure gauge. The zero mode in two form may seen to describe a topological torsion in a second order formalism. We work in a generalized geometry (\ref{gauge-22}) with $C=-(1/2)$. Alternately, an effective metric may be sourced by ${\tilde T}_{\mu\nu}$ in (\ref{gauge-31}) with $C=-(1/4)$). It is given by
\bea
&&ds^2=-\left (1-{{b^2}\over{r^2}} + {{b^2P^6}\over{r^8}} \right ) dt^2\ +\ 
\left ( 1 +{{b^2}\over{r^2}} - {{b^2P^6}\over{r^8}}\right ) dr^2\ +\ {{b^2}\over{r^2}}\left (1 - {{P^6}\over{r^6}} \right ) dt dr
\qquad\qquad {}\nonumber\\
&&\qquad\qquad\qquad\qquad\qquad\qquad\qquad +\ {{2bP^6}\over{r^6}}\left ( dt + dr \right )d\psi\
+\ \left ( 1 - {{P^6}\over{r^6}}\right ) r^2 d\Omega^2_3\ .\label{ads-3}
\eea
In the large $r$ limit, $i.e.$ $r$$>$$P$ and $r$$>$$b$ with a fixed $\alpha'$, on an effective $D_4$-brane, the geometry may be re-expressed as
\bea
ds^2=&-&\left (1-{{(2\pi\alpha' M)^2}\over{r^2}} + {{(2\pi\alpha')^2M^2P^6}\over{r^8}}\right ) dt^2\qquad\qquad {}\nonumber\\ 
&+& \left ( 1 -{{(2\pi\alpha' M)^2}\over{r^2}} + {{(2\pi\alpha')^2M^2P^6}\over{r^8}}\right )^{-1} dr^2\nonumber\\
&+& {{(2\pi\alpha' M)^2}\over{r^2}}\left ( 1 - {{P^6}\over{r^6}}\right ) dt dr + {{(4\pi\alpha')MP^6}\over{r^6}}\left ( dt + dr \right )d\psi 
\nonumber\\
&+&\left ( 1 - {{{P^6}}\over{r^6}}\right )  r^2 d\Omega^2_3 \ .\label{ads-4}
\eea
In the semi-classical regime, an effective $D_4$-brane geometry maps to a macroscopic black hole presumably described by an eleven dimensional $M$-theory. In fact, an empirical metric potential in (\ref{ads-4}) ensures the presence of $6$-extra compact dimensions transverse to an effective $D_4$-brane. A priori, the emergent black hole is characterized by two dynamical parameters $M=b/(2\pi\alpha')$  and a charge $P$. Nevertheless, the constants ($M,P, {\tilde P}$) in the frame-work may be identified with the extrinsic conserved quantities in a macroscopic charged black hole. 
A ${\tilde P}^3$ charge turns out to be topological. In principle, one may encounter a number of additional topological charges in the frame-work. A topological charge can modify an otherwise linear charge globally to its non-linear counterpart. It may seen to play a significant role 
underlying an angular momentum, an electric and a magnetic charges. It is indeed thought provoking to believe that the emergent black holes, underlying a geometric torsion, may be sourced by the conjectured dark energy in a fundamental gravity theory or in a $M$-theory. 

\sp
\noindent
Furthermore, the geometric torsion seems to break the $S^3$-spherical symmetry in the emergent black hole. However, the macroscopic charged black hole is asymptotically flat which would like to enforce ${\tilde\Lambda}=0$ in the frame-work. In the regime, we re-express the generic brane line-elements as:
\bea
&&ds^2=-\left (1-{{(2\pi\alpha' M)^2}\over{r^2}} \left ( 1 - {{P^6}\over{r^6}}\right )\right ) dt^2 + 
\left ( 1 -{{(2\pi\alpha' M)^2}\over{r^2}}\left ( 1 - {{P^6}\over{r^6}}\right ) \right )^{-1} dr^2\qquad {}\nonumber\\
&&\;\; + {{(2\pi\alpha' M)^2}\over{r^2}}\left ( 1 - {{P^6}\over{r^6}}\right ) dt dr 
+\ {{(4\pi\alpha')MP^6}\over{r^6}}\left ( dt + dr \right )d\psi +\left ( 1 - {{P^6}\over{r^6}}\right )  r^2 d\Omega^2_3 \ .\label{ads-5}
\eea
The rotating, charged, black hole is characterized by two physical horizons. A priori, two horizons may hint at two independent potentials defined respectively with two dynamical parameters ($M,P$). The black hole is characterized by an inner horizon at $r_-=(2\pi\alpha'M-\delta P)$ and an event horizon at $r_+=(P+\delta P)$. The angular velocity of the emergent black hole at its horizon $r_h$ becomes
\bea
\Omega_h&-&{{{\tilde G}_{t\psi}}\over{\tilde G}_{\psi\psi}}{\Big |}_{r_h} \ ,\nonumber\\
&=&+ {{(2\pi\alpha')}\over{r_h^2}}\left ( {{MP^6}\over{r_h^6-P^6}}\right )\ .\label{ads-51}
\eea
The semi-classical brane-world effective geometry leading to a charged black hole possesses a hypothetical reflection symmetry under $(t,r,M)\rightarrow (-t,-r,-M)$. The angular velocity on the brane-Universe changes $\Omega_h\rightarrow -\Omega_h$ under the symmetry. However within a gauge choice, the reflection symmetry is broken. The near horizon geometry on a mirror brane may be argued qualitatively from the brane (\ref{ads-5}) under $r\rightarrow -r$. The mirror black hole  presumably describes an anti $D_4$-brane (${\bar D_4}$-) in the frame-work. The effective ${\bar D}_4$-brane is also characterized by an inner and an event horizons. However these two brane-Universes are not connected by a causal patch, in a weak coupling regime, as the reflection symmetry is completely broken for a distant observer on a $D_4$-brane. In a strong coupling limit, the $(D{\bar D})_4$) may seen to form a bound state. The existence of an anti-brane is a new phenomenon in a frame-work. In fact, it is a consequence of the near horizon $D$-brane geometries which are solely sourced by a torsion. In other words, the angular momentum, incorporated by a torsion geometro-dynamics on an effective $D$-brane, plays a vital role to hint at the existence of its anti-brane. For instance, a system of a $D_4$-brane and ${\bar D}_4$-brane cancels the Ramond-Ramond (RR) charge of each other in type II superstring on $S^1$ to nucleate a $D_5$-brane. Needless to mention that the $D_5$-brane is sourced by a RR charge under a dual of two form. In principle, the nucleation of a higher dimensional $D_p$-brane from its lower dimensional $(D{\bar D})_{(p-1)}$ system saturates at $p=8$ in the formalism.

\subsection{Asymptotic AdS brane}
The difference in the metric potentials incorporate instabilities in the emergent black hole. The potentials would flow and balance among themselves to define an equipotential. However, restricting to a black hole geometry as viewed by a brane, the large rotating black hole tends to a stable vacuum and undergoes Hawking radiation. The torsion modes, in the regime, decouple through radiations from a macroscopic charged and rotating black hole. Its event horizon shrinks, angular velocity slows down and the inner horizon expands to arrive at an extremal limit $\delta P\rightarrow 0$. The extremal black hole may be defined by a horizon $r_h= (2\pi\alpha') M$. Then, the effective space-time on a $D_4$-brane reduces to the near horizon geometry of a five dimensional charged black hole. Under $r\rightarrow -r$, the near horizon geometry on a ${\bar D}_4$-brane may be obtained from that of a $D_4$-brane (\ref{ads-5}). Both the brane and anti-brane in its near horizon retain the spherical symmetry and reduce to their appropriate extremal black hole geometries. The are given by 
\be
ds^2=-\left (1-{{(2\pi\alpha' M)^2}\over{r^2}} \right ) dt^2 + 
\left ( 1 -{{(2\pi\alpha' M)^2}\over{r^2}}\right )^{-1} dr^2 \pm {{2(2\pi\alpha' M)^2}\over{r^2}} dt dr + r^2 d\Omega^2_3 \ .\label{ads-6}
\ee
The positive and negative signs in the emergent metric component ${\tilde G}_{tr}$, respectively, correspond to a $D_4$-brane and a ${\bar D}_4$-brane. 
However, the angular velocity (\ref{ads-51}) in the extremal limit reduces to a non-zero value. The angular velocities may be approximated to yield 
\be
\Omega_h^{ext}\rightarrow\ \pm {{J}\over{(2\pi\alpha') M}}\ ,\label{ads-61}
\ee
where $J>0$ is an infinitely large  dimension-less constant. Hence the extremal parameter $M$ may appropriately defines an angular momentum on an effective $D_4$-brane and its anti-brane. Their angular momenta are equal in magnitude but opposite in direction to each other. The rotating brane geometry is new and is a result of a geometric torsion in the frame-work. Furthermore, a non-zero ${\tilde G}_{tr}$ hints at the conserved 
charges $M$ and $-M$, respectively, on an effective $D_4$-brane and ${\bar D}_4$-brane. At the outset, the multiple role of $M$ imply
the presence of a topological charge ${\tilde P}$ on a brane-Universe. 

\sp
\noindent
Importantly, a contribution from a local torsion drops much faster than a topological torsion in an emergent geometry (\ref{ads-5}). Hence, an effective $D_4$-brane is more appropriately described by a non-propagating torsion in large $r$. In other words, a local torsion does not play a significant role in the near horizon $D_4$-brane geometry (\ref{ads-6}) constructed in the semi-classical regime. However, our analysis in Section-4 would like to unfold the mysteries of a local torsion to a $D_4$-brane in its small $r$ regime, leading to a primordial black hole. 
\begin{figure}
\centering
\mbox{\includegraphics[width=0.53\linewidth,height=0.31\textheight]{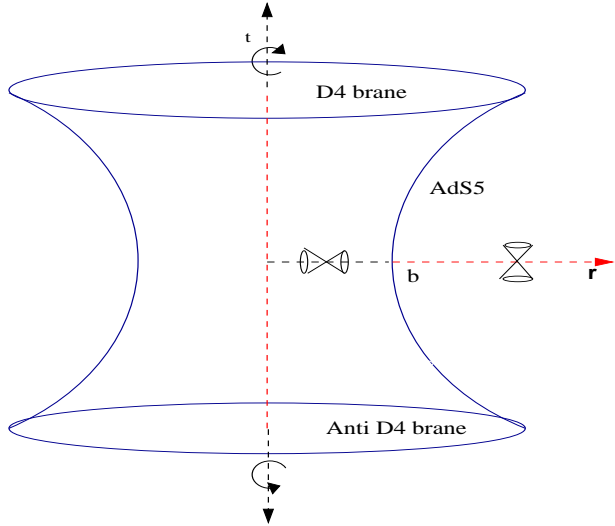}}
\caption{\it An effective $D_4$-brane, at large $r$, corresponds to an asymptotic AdS. A global mode $b$ is identified with an AdS radius of curvature. Within the horizon, the time-like coordinate becomes space-like and vice-versa. In the near horizon geometry, an effective anti $D_4$-brane may be obtained from an effective $D_4$-brane with $r\rightarrow -r$.}
\end{figure}
\noindent
On the other hand, the black hole in its extremal, limit $r\rightarrow r_h$, may further be explored with a subtlety for its near horizon brane $r_h=(2\pi\alpha')M - \epsilon$ and $\epsilon\rightarrow 0$. In the case, the horizon geometry may be approximated under an interchange within its longitudinal axes, $i.e.\ r\leftrightarrow t$. In fact, the interchange is enforced by a flip of its light-cone across the black hole horizon intuitively perceived by a distant observer. The near horizon rotating brane-world geometries, along an orthogonal radial axis to its original, yield: 
\be
ds^2 = - {{r^2}\over{b^2}}\ dt^2 + {{b^2}\over{r^2}}\ dr^2 \pm {{2b^2}\over{r^2}}dt dr\ +\ r^2 d\Omega^2_3\ . \label{ads-7}
\ee
The spherical symmetry is restored in the near horizon limit. The cross term in longitudinal spaces re-assure the fact that the stable non-trivial geometries are due to a conserved angular momentum primarily sourced by a global mode of a two form on a $D_4$-brane and ${\bar D}_4$-brane. At the first sight, an effective $D_4$-brane in large $r$ may formally be identified with an asymptotically flat rotating black hole (\ref{ads-6}). However, analysis for a near horizon $D$-brane and a ${\bar D}$-brane independently reveal an asymptotic AdS ``rotating'' geometry. From a global perspective, with a $D_4$-brane and ${\bar D}_4$-brane together in the frame-work, the charge/angular momentum cancel each other to define a typical asymptotic AdS. Then, the emergent geometry reduces to an elegant form:
\be
ds^2 = - {{r^2}\over{b^2}}\ dt^2 + {{b^2}\over{r^2}}\ dr^2 +\ r^2 d\Omega^2_3\ . \label{ads-8a}
\ee
At this point, we digress to illustrate the role of two form in the emergent $AdS_2\times S^3$ on a brane. Being a global mode, $b$ can not be gauged away in a gauge theory. It may intuitively describe a topological torsion in the frame-work. Thus a topological torsion may be argued to incorporate a cosmological constant into the Riemannian geometry. Though the black hole is asymptotically flat, its near horizon brane geometry approaches an asymptotic AdS. Thus in an effective curvature theory, a $D_4$-brane precisely describes an AdS space-time. In fact, the global mode $b$ is identified with the AdS radius of curvature. Hence, a two form may seen to source the vacuum energy density in an underlying effective curvature theory. Though a constant two form does not contribute to the local gauge dynamics on a $D_4$-brane, it can play a significant role to incorporate plausible local degrees in the Riemann curvature tensor theory. We recall that the cosmological constant becomes significant in Einstein's gravity, which is unlike to that in a gauge or quantum field theory. For instance, the presence of such a non-zero constant leads to a three dimensional charged black hole \cite{btz} which is otherwise not feasible. Though the irreducible conformal Weyl tensor becomes trivial in three dimensions, Riemann curvature tensor is  non-trivial in presence of a cosmological constant. Analysis may provide hints to resolve a disparity in local degrees between a two form gauge theory and an underlying Einstein's gravity in five dimensions.

\subsection{Cosmological constant}
A priori, the curvature scalar sourced by a dynamical torsion in an effective curvature theory on a $D_4$-brane is worked out to yield
\be
{\cal K}= -\ {{3P^6}\over{(4\pi\alpha') r^6}}\ .\label{ads-8}
\ee
The effective curvature blows up in the limit $r\rightarrow 0$. However, the curvature singularity is protected by the event horizon of an emergent black hole. In large $r$ limit, the curvature takes a small negative value for a fixed $P$. Alternately, the small value of the curvature on a $D_4$-brane in the regime may also be re-assured by the near horizon limit of the radiating macroscopic black hole. The gauge choice re-assures a small value of ${\tilde\Lambda}$ in the frame-work. The cosmological constant is worked out at the horizon $r_h$ of the extremal black hole, to yield 
\be
{1\over3}{\tilde\Lambda}_{\rm r_h} = {1\over{2\pi\alpha'}}\left (2 - {{P^6}\over{2b^6}}\right )\ .\label{ads-9}
\ee 
An emergent $AdS_2\times S^3$ space-time on an effective $D_4$-brane enforces ${\tilde\Lambda}_{r_h}<0$. It yields a relation between the dynamical parameters ($P,b$). It ensures $P^3>2b^3$, implying $P>b$, on a generic $D_4$-brane underlying an effective curvature frame-work. The fact $r>P>b$ in the semi-classical regime on an effective $D_4$-brane is in conformity with its inner and event horizon potentials, respectively, being tuned by the parameters $b$ and $P$. In other words, the near horizon geometry of an emergent Schwarzschild black hole corresponds to an effective $D_4$-brane which is identified with an asymptotic AdS. Thus a $D_4$-brane may be viewed as an AdS-brane in the frame-work. Interestingly, a class of supersymmetric $AdS_2\times S^2$ geometry on a $D_3$-brane has been discused in ref.\cite{bachas} in a different context.

\section{Quantum regime: Discrete torsion on ${\mathbf{D_4}}$-brane}
We revisit the effective geometries (\ref{ads-3}) in the small $r$ limit to explore certain aspects of quantum geometric phases in the theory. Importantly, the quantum brane-world effective geometries are perceived under appropriate Weyl scaling(s) of a generalized metric. The geometry in the limit approaches a small primordial black hole at Planck scale. In the limit, the string coupling $g_s$$>>$$1$. Thus a $D_4$-brane becomes light and is dominantly governed by quantum physics. The existence of an anti $D_4$-brane effective geometry, as discussed in section 3.1, is revisited in the quantum domain to explore a bound state of $(D{\bar D})_4$. As a result, the net effect of angular momentum vanishes in a global geometry consisting of a $D_4$-brane and a ${\bar D}_4$-brane. The fragmented geometries describe an emergent dS leading to a rotating charged black hole. We analyze the dS geometric patches to address the Big Bang at the beginning of a brane-Universe in the frame-work.
\subsection{de Sitter causal patches}
The emergent quantum geometry, in the regime, is precisely governed by a dynamical torsion in the frame-work. In the case, we define the small $r$ limit by incorporating appropriate dynamical bounds on $r$. In particular, the quantum regime on an effective $D_4$-brane is defined by $P<r<b$, with $P\neq 0$, while $\alpha'$ is kept fixed. It may seen to introduce bounds on the effective potentials on a $D_4$-brane. They are given by
\be
{{P^6}\over{r^6}}<<{{r^2}\over{b^2}}<<1\ .\label{dS-0}
\ee 
The bounds may allow one to view the brane-world geometry (\ref{ads-3}) as a small black hole. 
In fact the quantum geometry is derived under an explicit Weyl scaling:
\be
{\tilde G}_{\mu\nu}= {{b^2}\over{r^2}} {{\tilde G}'}_{\mu\nu}\ .\qquad\qquad {}\label{dS-1}
\ee
A large conformal factor due to the allowed range of $r$ enforces a quantum geometry underlying a transformed metric ${{\tilde G}'}_{\mu\nu}$ and hence the small $r$ limit. The arbitrary-ness in the constant $C$ is exploited in the effective metric ${\tilde G}_{\mu\nu}$ or in its source ${\tilde T}_{\mu\nu}$ by considering $C=(1/2)$ in ${\tilde G}_{\mu\nu}$ or $C=(3/4)$ in ${\tilde T}_{\mu\nu}$ in addition to the effective geometry in eq.(\ref{ads-3}). The effective curvature on a $D_4$-brane is worked out, a priori, with a transformed metric. The effective $D_4$-brane geometries in the quantum regime becomes
\bea
ds^2= &-&\left (-\left (1-{{r^2}\over{b^2}}\right ) \mp {{P^6}\over{r^6}}\right ) dt^2 + 
\left ( \left (1 -{{r^2}\over{b^2}}\right ) \mp {{P^6}\over{r^6}}\right )^{-1} dr^2 + {{2r}\over{b}}\left ( dt + dr \right ) r d\psi\nonumber\\
&+& \left ( 1 \pm {{P^6}\over{r^6}}\right ) \left ( 2 dt dr - {{2r}\over{b}}\left ( dt + dr \right ) r d\psi + {{r^4}\over{b^2}}  d\Omega^2_3 \right )\ .\label{dS-2}
\eea
The constants $b$ and $P$ have been identified, respectively, with a cosmological scale and an energy scale leading to dS geometries.
Since the causal properties remain unchanged under a Weyl scaling, the scaled down transformed geometries enclose a curvature singularity (\ref{ads-8}) at $r\rightarrow 0$. The presence of a curvature singularity, at $r\rightarrow 0$, makes the discrete torsion ($P\neq 0$) significant in the regime.  The cosmological constant (\ref{ads-9}) for an effective $D_4$-brane vacuum (\ref{dS-2}) satisfies ${\tilde\Lambda}>0$ at its horizon(s). The small value of ${\tilde\Lambda}$ may also be argued in quantum geometry where the stringy ($\alpha'$)-corrections are significant. Thus an effective $D_4$-brane, in small $r$, may describe a dS vacuum. 
\subsection{Big Bang: Branes within branes}
The quantum geometries (\ref{dS-2}) underlying a patch of dS may simply be re-expressed as:
\bea
&&ds^2= \left (1-{{r^2}\over{b^2}} \pm {{P^6}\over{r^6}}\right ) dt^2 + 
\left ( 1 -{{r^2}\over{b^2}} \mp {{P^6}\over{r^6}}\right )^{-1} dr^2 + 2\left ( 1 \pm {{P^6}\over{r^6}}\right ) dt dr\nonumber\\
&&\qquad\qquad\qquad\qquad\qquad\qquad\qquad \mp\ {{2P^6}\over{br^4}}\left ( dt + dr \right )d\psi +\left ({{r^2}\over{b^2}}\pm {{{P^6}}\over{b^2r^4}}\right ) r^2 d\Omega^2_3 \ .\label{dS-3}
\eea
An euclidean time, presumably underlying a thermal description for dS geometry, is noteworthy in the frame-work. In fact, a Weyl scaling of the effective metric (\ref{ads-3}) seems to have introduced a change in metric signature. Apparently, the metric scaling in the regime, has forced time to be on $S^1$. It hints at a notion of temperature naturally in the dS vacua in the regime $P<r<b$ on an effective $D_4$-brane and its anti-brane. The quantum black hole may further be explored for an enhanced understanding of an effective $D_4$-brane geometries at small $r$. For simplicity, we re-write the line-element(s) 
as 
\bea
&&ds^2= \left (1-{{r^2}\over{b^2}} \right ) dt^2\ +\  
\left ( 1 -{{r^2}\over{b^2}}\right )^{-1} dr^2\ +\ 2 dt\ dr\ +\ {{r^4}\over{b^2}} d\Omega^2_3\qquad\qquad\qquad\qquad\qquad\nonumber\\ 
&&\quad\quad\qquad\qquad\qquad\qquad\pm\ {{P^6}\over{b^2r^4}}\left ( {{b^2}\over{r^2}} (dt +dr)^2\ -\ 2b\ (dt + dr)d\psi\ +\ r^2 d\Omega_3^2\right )\ .\label{dS-4}
\eea
The effective brane geometries are splitted and they are explicitly expressed as a pure dS vacuum, in static coordinates, with a torsion fluctuation. In a limit $P\rightarrow 0,\ i.e.$ in absence of a quantum fluctuation, the effective metric underlying an euclidean pure dS turns out to be singular. It re-assures the significant role played by a quantum fluctuation underlying a geometric torsion on a $D_4$-brane in its small $r$ regime.
Interestingly a degenerate metric, in absence of a quantum fluctuation, hints at a Big Bang at the beginning of a brane-Universe. Presumably an Universe began with a Big Bang in the near (cosmological) horizon ($r_c\pm\epsilon$) with an euclidean time underlying an non-perturbative frame-work. The quantum fluctuations, with $P\neq 0$, may thought to be sourced by the dark energy in the effective curvature theory. 

\sp
\noindent
Under $r\rightarrow -r$, a $D_4$-brane effective geometry with a left handed angular momentum may seen to generate an anti $D_4$-brane vacuum with a right handed angular momentum and vice-versa. The ${\bar D}_4$- and $D_4$- propagate, respectively, along $-r$ and $+r$ in the near (cosmological) horizon. Hence their geometries would differ only by a sign in the $G_{tr}$ component (\ref{dS-2}). The ${\bar D}_4$-brane geometry may be given by
\bea
&&ds^2= \left (1-{{r^2}\over{b^2}} \right ) dt^2\ +\  
\left ( 1 -{{r^2}\over{b^2}}\right )^{-1} dr^2\ -\ 2 dt\ dr\ +\ {{r^4}\over{b^2}} d\Omega^2_3\qquad\qquad\qquad\qquad\qquad\nonumber\\ 
&&\quad\quad\qquad\qquad\qquad\qquad\pm\ {{P^6}\over{b^2r^4}}\left ( {{b^2}\over{r^2}} (dt -dr)^2\ -\ 2b\ (dt - dr)d\psi\ +\ r^2 d\Omega_3^2\right )\ .\label{dS-5}
\eea
The emerging notion of an effective ${\bar D}_4$-brane in the near (cosmological) horizon is analogous to that discused in a semi-classical regime. However, a $(D{\bar D})_4$-brane pair in the regime forms a bound state unlike to that in a semi-classical regime. The bound state signifies a strong (string) coupling $g_s>>1$ and hence is in agreement with a non-perturbative metric. Considering both $D_4$-brane and ${\bar D}_4$-brane together, $i.e.$ from a global perspective, the charges associated with the metric components ${{\tilde G}'}_{rt}$ and ${{\tilde G}'}_{r\psi}$ annihilate each other. The remaining cross metric component ${{\tilde G}'}_{t\psi}$ naively signify a non-zero angular momentum. The effective global geometry is given by
\bea
&&ds^2= \left (1-{{r^2}\over{b^2}} \pm {{P^6}\over{r^6}} \right ) dt^2\ +\  
\left ( 1 -{{r^2}\over{b^2}} \mp {{P^6}\over{r^6}}\right )^{-1} dr^2\qquad\qquad {}\nonumber\\
&&\qquad\qquad\qquad\qquad\qquad\qquad\mp {{2P^6}\over{br^4}} dtd\psi\ +\ \left ({{r^2}\over{b^2}} \pm {{P^6}\over{b^2r^4}}\right )\ r^2 d\Omega^2_3
\ .\label{dS-5ab}
\eea
An analytic continuation to real time, ensures a small but imaginary angular momentum ${\tilde J}=i(P/b)^6$ at the cosmological horizon in a global scenario. However in absence of a geometric torsion, the near (cosmological) horizon brane geometry describes a pure dS with a real time. It may imply
that the (dark) energy presumably coupled to an euclidean pure dS at the cosmological horizon in a Big Bang. The notion of time becomes significant on a brane and its anti-brane created at the Big Bang. Interestingly, the metric singularity in absence of torsion 
independently on a brane and an anti-brane is absent in a global scenario. However, the singularity is re-instated in the dS curvature tensors. 
In other words, a pure dS  with a non-zero small $P$ at the past horizon may be identified with a Big Bang, $i.e.$ the creation of a brane (anti-brane) pair in the formalism. Under a change $r\rightarrow -r$, the $G_{tr}$ and $G_{r\psi}$ components change their sign in the line-element (\ref{dS-3})) and the emergent dS patch would presumably describe a Big Crunch underlying an annihilation of a brane world or creation of an anti-brane. Interestingly, $D$-brane in a Big Bang has been studied in a different  context in ref.\cite{hikida-nayak-panigrahi}.

\sp
\noindent
Now, we revisit a non-perturbative (quantum) correction in eq.(\ref{dS-4}) to a pure $dS_5$ rotating geometry, underlying an effective $D_4$-brane, in light-cone coordinates $x_{\pm}=(t\pm r)$. It is given by
\bea
ds^2_{q+}&=&f^{nz}_{\mu\nu}\ dx^{\mu}dx^{\nu}\nonumber\\
&=&\pm\ {{P^6}\over{b^2r^4}}\left ( {{b^2}\over{r^2}} dx_{+}^2\ -\ 2b\ dx_{+}d\psi\ +\ r^2 d\Omega_3^2\right )\ .\label{dS-4a}
\eea
Similarly the geometric correction, underlying an effective ${\bar D}_4$-brane, may be expressed as
\be
ds^2_{q-}=\pm\ {{P^6}\over{b^2r^4}}\left ( {{b^2}\over{r^2}} dx_{-}^2\ -\ 2b\ dx_{-}d\psi\ +\ r^2 d\Omega_3^2\right )\ .\label{dS-5a}
\ee
Firstly, a quantum fluctuation leading to an effective $D_4$-brane and a ${\bar D}_4$-brane are manifestations of a geometric torsion ($P\neq 0$) dynamics in the frame-work. They are significantly large in the regime. A priori, a fluctuation does not seem to elope with time, which is in agreement with a thermal notion inherent with a hot dS vacuum. Secondly, a fluctuation on an effective $D_4$-brane is independent to that on its anti-brane. For instance, the radial coordinates $x_+$ and $x_-$ are independent of each other. Most importantly, the four dimensional non-perturbative fluctuation to a
five dimensional pure dS is remarkable. With a notion of euclidean time, the patches may be re-expressed as:
\be
ds^2_{q\pm}\rightarrow \left ( dt^2_e\ +\ {{b^2}\over{r^2}} dx_{\pm}^2\ -\ {{2b}\over{r}}\ dt_e dx_{\pm}\ +\ r^2 d\Omega_2^2\right )\ .\label{dS-ab}
\ee
At the past horizon $r\rightarrow r_c=b$, the line-element drastically simplifies to yield
\be
ds^2_{q\pm}\rightarrow \left ( d\rho_{\pm}^2\ +\ b^2 d\Omega_2^2\right )\ ,\label{dS-ba}
\ee
where $\rho_{\pm}= (t_e -x_{\pm})$ correspond to new radial coordinates, respectively, on an effective $D_4$-brane and an ${\bar D}_4$-brane. The $S^2$ geometry decouples from the radial coordinate in the near (cosmological) horizon geometry. It further shows that the near horizon causal patches become flat. They ensures a thermal phase described by three spatial coordinates with $R\times S^2$ topology.

\sp
\noindent
It is thought provoking to believe that a brane-Universe was created at a Big Bang from a degenerate metric underlying an emergent
(euclidean) pure de Sitter. The quantum geometry was hot and may solely be characterized by an angular momentum $b$, which in turn may be identified with a radius at the cosmological horizon $r_c=b$. Instantaneously the dS becomes non-degenerate due to the dark energy sourced by a geometric torsion $P$ in the formalism. Presumably upon time, the brane-world Universe has cooled down to yield a non-trivial space-time underlying an emergent dS quantum black hole. Thus, an effective $D_4$-brane may be understood from the polarization of a vacuum, at the cosmological horizon, by the dark energy sourced by a discrete torsion in the frame-work. With a Big Bang, or equivalently with a Big Crunch, the vacuum is polarized to yield a pair of effective $D_4$-brane and an anti $D_4$-brane, which move respectively along $r$ and $-r$ in the near (cosmological) horizon geometry. 

\sp
\noindent
Interestingly, the origin of an effective $D_4$-brane Universe may be argued all the way down from an effective  $D$-instanton due to the significant role played by the non-zero modes of NS-NS two form in the string bulk. At the Big Bang, a discrete torsion becomes significant and non-perturbatively incorporates a geometric fluctuation to define an instantaneous point, $i.e.$ a $D$-instanton. Its subsequent growth with dark energy, gave birth to a $D$-particle. The world-line of a $D$-particle may be described by $\rho_+$ in eq.(\ref{dS-ba}). Alternately, an increase in $P$, may be viewed as an annihilation of a pair of $D$-particle and its anti particle to yield an effective $D$-string. As the fluctuations grow, a $D$-string on $S^2$ along with a $D$-particle in its bulk describe a $D$-membrane. The quantum geometry associates it-self with an euclidean time $t_e$ in the near horizon, which in turn describes a four dimensional geometric fluctuations on a $D_3$-brane. Similarly, the origin of an effective ${\bar D}_4$-brane may be argued from a ${\bar D}$-instanton, via a ${\bar D}$-string and a ${\bar D}$-membrane, to a ${\bar D}_3$-brane. In other words, an effective $D_4$-brane vacuum in the quantum regime, is polarized in presence of a light torsion to yield a pair of $(D{\bar D})_3$-branes. With an increase in dark energy, an effective $D_3$-brane vacuum is polarized further to yield a pair of $(D{\bar D})_2$-branes. The geometric transitions in steps continue with an increase in energy and lead to $D_2\rightarrow (D{\bar D})_1$. It is followed by $D_1\rightarrow (D{\bar D})_0$ and $D_0\rightarrow (D{\bar D})_{-1}$. Intuitively, the geometric evolutions of space-time began with a Big Bang (Big Crunch) in an effective $D_4$-brane (${\bar D}_4$-brane) in a quantum regime, which are associated with a series of nucleation of lower branes and its anti-branes. Arguably, an effective $D_4({\bar D}_4)$-brane may be connected to the Big Bang (Crunch) through the lower barnes within them: 
\bea
&&{\rm (Big\ Bang)}\longrightarrow D_{-1}\longrightarrow D_0\longrightarrow D_2\longrightarrow D_3\longrightarrow D_4\nonumber\\
{\rm and}\;\ &&{\rm (Big\ Crunch)}\longrightarrow {\bar D}_{-1}\longrightarrow {\bar D}_0\longrightarrow {\bar D}_2\longrightarrow {\bar D}_3\longrightarrow {\bar D}_4
\ .\label{dS-abc}
\eea
It implies that an effective $D_4$-brane (or ${\bar D}_4$-) Universe, in the frame-work, contains all of its lower $D$-branes (or ${\bar D}$-). 
In fact, a geometric torsion theory on an effective $D_p$-brane allows a generalized description of branes within branes \cite{douglas-cargese} in a $D=10$ type II superstrings on $S^1$. A typical $D_p$-brane is described by large density of ($p-2$)-branes. Furthermore, each $(p-2)$-brane world-volume contains a large density of ($p-4$)-branes, and so on, in either type IIA or IIB superstring theories in $D=10$. The difference of two spatial dimensions, between a brane and its brane within, is essentially due to the zero modes of NS-NS two form which couples appropriately to Ramond-Ramond (RR) form fields in Chern-Simons action underlying a $D_p$-brane in type II superstring theories. However, the generalized notion of lower branes within a higher brane is specific to an effective curvature theory. In addition to a zero mode on a $D_4$-brane, it takes into account a non-zero mode of NS-NS two form in a higher dimensions with compact dimensions. The potential in the effective metric further ensures the presence of extra compact dimensions transverse to a $D_5$-brane for small $r$.

\subsection{Painleve de Sitter vacuum}
Now, we analyze the emergent $dS_5$ patches on a near horizon $D_4$-brane for a Painleve vacuum. The dS brane (\ref{dS-2}) simplifies drastically, in light cone coordinates to yield an $S^2$-symmetric geometry. It is given by
\be
ds^2= -{{r^2}\over{b^2}}\ dx_{+} \left ( dx_{-} - 2b\ d\psi \right)\ +\ \left ( 1 \pm {{P^6}\over{r^6}}\right ) 
\left ( \left (dx_{+} - {{r^2}\over{b}}\ d\psi\right )^2 + {{r^4}\over{b^2}}\  d\Omega^2_2 \right )\ .\label{dS-6}
\ee 
The passage through a coordinate singularity at $r\rightarrow r_c=b$, in a pure dS patch, has been eased out in light cone coordinates \cite{parikh-wilczek,parikh,medved}. The Painleve coordinates are known to play a significant role to describe the geometry of a slowly evaporating black hole. Though, a quantum fluctuation played a vital role at the cosmological horizon, its extremely small value at the creation of the quantum brane-world may be ignored. In the approximation, the geometries converge to a rotating five dimensional, tiny, single brane-Universe at the near (cosmological) horizon. A priori, the line-element becomes
\be
ds^2= -\ dx_{+} dx_{-}\ +\ 2r\ dx_+ d\psi\ +\ d{\tilde x}^2\ +\ r^2\ d\Omega^2_2\ ,\label{dS-7}
\ee
where $d{\tilde x}= (dx_{+} - r d\psi)$. The subtle geometrical patch, on a quantum $D_4$-brane in the near (cosmological) horizon, may further be worked out to yield
\be
ds^2=\ 2dr^2 + 2 dr dt + r^2\ d\Omega^2_3\ .\label{dS-8}
\ee
Alternately, the emergent dS brane geometries (\ref{dS-2}) in its near (cosmological) horizon may as well be approximated to yield
\be
ds^2=\ \mp {{b^6}\over{P^6}} dr^2 - 2 dt dr + r^2\ d\Omega^2_3\ .\label{dS-9}
\ee
The presence of a constant large scale $(b^6/p^6)>>1$ in ${\tilde G}_{rr}$ further re-assures a small scale in the emergent geometry (\ref{dS-9}). Apparently, the reduced geometrical patches, at the cosmological horizon, provoke two thoughts in order. Firstly, it hints at a decoupling of the $S^3$-symmetric transverse space from its longitudinal geometric patch at the creation of a brane pair. In fact, the spherically symmetric space at Big Bang (or Big Crunch) may have been governed by a fixed spatial radius $b$, the maximum value of $r$ in the frame-work. The longitudinal geometry collapses at the cosmological horizon and the effective geometry may completely be described by a spherically symmetric three dimensional space. Secondly, the geometrical patch may be identified with a pure dS vacuum in static coordinates. At this point, we digress to a pure dS static vacuum in Painleve coordinates. It is given by
\be
ds^2= -\left ( 1 -{{r^2}\over{b^2}}\right ) dt^2\ {\pm} 2{{r\over{b}}}\ dt dr\ +\ dr^2\ +\ r^2\ d\Omega^2_3\ .\label{dS-10}
\ee
Importantly, the reduced geometries (\ref{dS-8}) and (\ref{dS-9}) describe causal patches of a Painleve dS vacuum, at the cosmological horizon, in five dimensions. Arguably, the effective geometry has began to build up on a quantum $D_4$-brane at the past horizon. A discrete dynamical torsion, in disguise of dark energy in an emergent gravity scenario, plays a significant role to source a dynamical gravity.

\subsection{de Sitter tunneling vacua}
Now, we focus on some of the effective $D_4$-brane geometries leading to a primordial dS black hole. In particular, the energy is quantized with a light quantum in the regime, which in turn signify a propagating torsion in the frame-work. An effective $D_4$-brane, with a geometric torsion, is a new phenomenon and may need further attention. A discrete torsion \cite{douglas} in the formalism may be exploited for its intrinsic angular momentum, which in turn may allow one to perform a discrete transformation on the mixed dS geometric patches (\ref{dS-2}). The transformation separates out a SdS 
black hole from a TdS. We analyze the Hawking radiation in a SdS black hole leading to a meta-stable Nariai black hole in the quantum regime. Interestingly, the near horizon geometry in an emergent Nariai black hole is identified with a patch of TdS black hole.

\sp
\noindent
In the context, an effective $D_4$-brane in the quantum regime leading to the emergent dS geometries (\ref{dS-2}) are naively described by eight horizons. However, two of them are physical (real and positive). They are: an event horizon at $r_e=(P+\delta P)$ and a cosmological horizon at $r_c=(b-\delta P)$, where $\delta P$ denotes a quantum fluctuation. An observer in both the geometries may seen to be located within the horizons, which is in agreement with $P<r<b$. In addition, the black hole geometries, on a effective $D_4$-brane, are characterized by an angular momentum, which has been argued to be nullified by its anti-brane in global geometry. The maximal dS black holes presumably approach a meta-stable vacuum where its horizon areas are equal and the spatial quantum phase disappears. Nevertheless, the horizons are separated by an euclidean time-like geometric phase $\delta P\rightarrow (\delta P)_t\neq 0$, which defines the allowed regime for an observer in SdS.

\sp
\noindent
On the other hand, the emergent dS geometries (\ref{dS-2}) neither correspond to a typical SdS  nor to a TdS even within the purview of euclidean longitudinal space-time. However, they may be viewed as an orthogonal combination of a Schwarschild and a topological metric components (${{\tilde G}'}_{tt}$ ${{\tilde G}'}_{rr}$) with an euclidean signature. To address the SdS and TdS black holes underlying an effective $D_4$-brane, we define a generalized ($2\times2$) matrix $M$ containing the longitudinal components of the metric quanta with euclidean signature. The cross terms in eq.(\ref{dS-3}) may be argued to nullify due to the existence of an ${\bar D}_4$-brane. A generalized matrix may be constructed for the longitudinal matrix components. It is given by
\begin{equation}
M=\frac{1}{2}\left( \begin{array}{ccc}
{{\tilde{G}}'}_{tt}(S)& & {{\tilde{G}}'}_{rr}(S)\\
 & & \\
{{\tilde{G}}'}_{rr}(T) & & {{\tilde{G}}'}_{tt}(T)
\end{array} \right)\ . \label{T-1}
\end{equation}

\bea
{\rm where}\; && {{\tilde G}'}_{tt}(S)= \left (1-{{r^2}\over{b^2}} - {{P^6}\over{r^6}}\right )\ ,\qquad\quad{{\tilde G}'}_{rr}(S)= \left (1-{{r^2}\over{b^2}} - {{P^6}\over{r^6}}\right )^{-1}\ ,\qquad\qquad {}\nonumber\\
&& {{\tilde G}'}_{tt}(T)= \left (1-{{r^2}\over{b^2}} + {{P^6}\over{r^6}}\right )\quad {\rm and}\quad\;
{{\tilde G}'}_{rr}(T)= \left (1-{{r^2}\over{b^2}} + {{P^6}\over{r^6}}\right )^{-1}\ .\label{T-2}
\eea
The metric components in ${{\tilde G}'}_{\mu\nu}(S)$ and ${{\tilde G}'}_{\mu\nu}(T)$, respectively, describe a SdS 
and a TdS on an effective $D_4$-brane. The inverse matrix becomes 
\begin{equation}
M^{-1}=\frac{1}{2\det M }\left( \begin{array}{ccc}
{{\tilde{G}}'}_{tt}(T) & & -{{\tilde{G}}'}_{rr}(S)\\
 & & \\
-{{\tilde{G}}'}_{rr}(T)& &{{\tilde{G}}'}_{tt}(S)
\end{array} \right)\ . \label{T-3}
\end{equation}

\vspace{.2in}
\begin{equation}
{\rm Then}\qquad\qquad M\left( \begin{array}{c}
1\\
\\
0
\end{array}\right)=\frac{1}{2}\left( \begin{array}{c}
{{\tilde{G}}'}_{tt}(S)\\
\\
{{\tilde{G}}'}_{rr}(T)
\end{array}\right) 
\qquad{\rm and} \qquad M\left( \begin{array}{c}
0\\
\\
1
\end{array}\right)=\frac{1}{2}\left( \begin{array}{c}
{{\tilde{G}}'}_{rr}(S)\\
\\
{{\tilde{G}}'}_{tt}(T)
\end{array}\right) \ . \label{T-3a}
\end{equation}

\vspace{.1in}
\noindent
The quantum patches, in the longitudinal space, are a mixture of the SdS and the TdS patches. 
They have already been obtained in eq.(\ref{dS-3}). Interestingly, the ($\det M$) may be computed to yield
\be
\det M= -{{r^2}\over{b^2}}\ .\label{T-4}
\ee
The determinant at the cosmological horizon ensures ($\det M=-1$) a discrete transformation, underlying a generalized non-degenerate matrix $M$ for its projection on two dimensional column vectors in longitudinal space. Interestingly under an inverse matrix operation, we obtain the appropriate longitudinal metric quanta leading to a SdS and a TdS.They are given by
\begin{equation}
 M^{-1}\left( \begin{array}{c}
1\\
\\
0
\end{array}\right)=\frac{1}{2}\left( \begin{array}{c}
-{{\tilde{G}}'}_{tt}(T)\\
\\
{{\tilde{G}}'}_{rr}(T)
\end{array}\right) 
\qquad{\rm and} \qquad M^{-1}\left( \begin{array}{c}
0\\
\\
1
\end{array}\right)=\frac{1}{2}\left( \begin{array}{c}
{{\tilde{G}}'}_{rr}(S)\\
\\
-{{\tilde{G}}'}_{tt}(S)
\end{array}\right) \ . \label{T-5}
\end{equation}

\vspace{.1in}
\noindent
It implies that the effective longitudinal metric components in eq.(\ref{dS-3}) on a $D_4$-brane, under an improper transformation, can lead to a SdS and a TdS. Importantly, the discrete projection by an inverse generalized matrix re-establishes the lorentzian signatures. Then, the $D_4$-brane effective geometry corresponding to a SdS is given by
\bea
&&ds^2= -\left (1-{{r^2}\over{b^2}} - {{P^6}\over{r^6}}\right ) dt^2 + 
\left ( 1 -{{r^2}\over{b^2}} - {{P^6}\over{r^6}}\right )^{-1} dr^2 + 2\left ( 1 - {{P^6}\over{r^6}}\right ) dt dr\nonumber\\
&&\qquad\qquad\qquad\qquad\qquad\qquad\; +\ {{2P^6}\over{br^4}}\left ( dt + dr \right )d\psi +\left ({{r^2}\over{b^2}}- {{{P^6}}\over{b^2r^4}}\right ) 
r^2 d\Omega^2_3 \ .\label{T-6}
\eea
The SdS is defined with a negative gravitional mass and an observer is restricted to a temporal phase between the cosmological horizon $r_c$ and 
the event horizon $r_e$. On the other hand, the effective $D_4$-brane geometry corresponding to a TdS is given by
\bea
&&ds^2= -\left (1-{{r^2}\over{b^2}} + {{P^6}\over{r^6}}\right ) dt^2 + 
\left ( 1 -{{r^2}\over{b^2}} + {{P^6}\over{r^6}}\right )^{-1} dr^2 + 2\left ( 1 + {{P^6}\over{r^6}}\right ) dt dr\nonumber\\
&&\qquad\qquad\qquad\qquad\qquad\qquad\; -\ {{2P^6}\over{br^4}}\left ( dt + dr \right )d\psi +\left ({{r^2}\over{b^2}}+ {{{P^6}}\over{b^2r^4}}\right ) 
r^2 d\Omega^2_3 \ .\label{T-7}
\eea
Unlike a SdS, a TdS is defined with a positive gravitational mass and an observer is at the otherside of its cosmological horizon. 
Interestingly, the computation of angular velocity in SdS and TdS at their respective cosmological horizons show that 
\be
\Omega_c^{\rm SdS}=-\Omega_c^{\rm TdS}=\ {{P^6}\over{b^7}}\ .\label{T-71}
\ee
It is evident that both the emergent black holes (SdS and TdS) are rotating with a small angular velocity in opposite direction to each other at their cosmological horizons. The angular velocity of SdS at the event horizon is computed to yield
\be
\Omega_e^{\rm SdS}=\ J{{b}\over{P^2}}\ .\label{T-72}
\ee
Thus an emergent SdS is rotating  with a very large angular velocity at its event horizon $r_e$. At this juncture, we recall that a pure dS at the Big Bang was primarily sourced by a coupling of a zero mode to a topological torsion in an effective curvature formulation underlying a $D$-brane in a type II super-string theory. A torsion coupling to a pure dS presumably began at the cosmological horizon, which gave birth to a slowly rotating brane (and anti brane) in the near (cosmological) horizon. An increase in torsion, naturally increases the angular momentum of the brane-pairs at their event horizon(s). The difference in angular momentum between an event horizon and a cosmological horizon causes instability in an emergent SdS black hole, which in turn intiates Hawking radiations. Thus, an evaporation of SdS black hole, followed by a fragmentation of dS, leads to a Nariai geometry. From a global perspective, with an effective $D_4$-brane and ${\bar D}_4$-brane together, the emergent SdS and TdS are respectively given by 
\bea
&&ds^2= -\left (1-{{r^2}\over{b^2}} - {{P^6}\over{r^6}}\right ) dt^2 + 
\left ( 1 -{{r^2}\over{b^2}} - {{P^6}\over{r^6}}\right )^{-1} dr^2\qquad\qquad\qquad {}\nonumber\\
&&\qquad\qquad\qquad\qquad\qquad\qquad\quad +\ {{2P^6}\over{br^4}} dt d\psi +\left ({{r^2}\over{b^2}}- {{{P^6}}\over{b^2r^4}}\right ) r^2 d\Omega^2_3 \label{T-61}
\eea
\vspace{-.1in}
and
\vspace{-.05in}
\bea
&&ds^2= -\left (1-{{r^2}\over{b^2}} + {{P^6}\over{r^6}}\right ) dt^2 + 
\left ( 1 -{{r^2}\over{b^2}} + {{P^6}\over{r^6}}\right )^{-1} dr^2\qquad\qquad\qquad\nonumber\\ 
&&\qquad\qquad\qquad\qquad\qquad\qquad\quad -\ {{2P^6}\over{br^4}} dt d\psi+\left ({{r^2}\over{b^2}}+ {{{P^6}}\over{b^2r^4}}\right ) r^2 d\Omega^2_3 \ .\label{T-71a}
\eea
On the other hand, the effective $D_4$-brane geometries differ significantly, in their potential, from the five dimensional dS geometries \cite{cai-myung-zhang,rong-cai,medved,kanti} in Einstein's theory. In an effective curvature prescription, the source potentials underlying a SdS and a TdS hint at a nine dimensional effective space-time at Planck scale on a $D_4$-brane. The extra dimensions are transverse to a $D_4$-brane in a ten dimensional dimensional type II superstrings on $S^1$. In fact, the emergent dS black holes on an effective $D_4$-brane are primarily described by a geometro-dynamics of a torsion in a second order formalism. In other words, the significant role played by the non-zero mode in the string bulk on an effective $D$-brane is re-assured in the frame-work.

\sp
\noindent
The higher dimensional emergent SdS (\ref{T-6}), on an effective $D_4$-brane, is described by a cosmological horizon at $r_c=(b-\delta P)$ and an event horizon at $r_e=(P+\delta P)$. The potential difference, between the two horizons introduce instabilities into a SdS geometry. As a result, the black hole undergoes Hawking radiation. Its event horizon expands to $r_e\rightarrow {\hat r}_e=(P+n\delta P)$ and its cosmological horizon apparently shrinks to $r_c\rightarrow {\hat r}_c=(b-m\delta P)$, where ($n,m$) take integer values. The expansion of the event horizon may be interpreted due to the growth of the dynamical constant $P$ in the formalism. It may also be linked to the growth of dark energy in an effective metric theory. Most importantly, an increase in event horizon is associated with a decrease in cosmological horizon. With a subtlety, it implies that an increase in $P$ causes $b$ to 
decrease. The variation in the dynamical constants, $i.e.$ with a set ($P_i\leftrightarrow 1/b_i$), are evident due to a metric fluctuation in the frame-work, which in turn lead to a conservation of total energy in space-time.

\sp
\noindent
In the Nariai limit, though the area of the horizons tend to equal, they are separated by a temporal geometric phase 
$(\delta P)_t\rightarrow |{\hat r}_c-{\hat r}_e|_t\neq 0$ between the equipotential. In fact, the interpolating potential between the two horizons may be worked out to yield a global maximum in the geometric phase defined by a temporal $r$ in the regime.  In particular, the equipoential with a spatial $r$ may be viewed at ${\hat r}_e\rightarrow {\hat r}_c$, but for an observer, $i.e.$ along a temporal $r$, a non-zero geometric phase $(\delta P)_t$ separates the two horizons ${\hat r}_e$ and ${\hat r}_c$ with a gobal maximum. In fact, the potential starts to increase from $(P+n\delta P)=P_{max}$ to a maximum and then falls to arrive at an equipotential at $(b-m\delta P)=b_{min}$. Intuitively the maximum, along a temporal $r$-coordinate, may be interpreted as a shock wave peak along a spatial $r$-coordinate at $r\rightarrow {\hat r}_e\rightarrow {\hat r}_c$ in a SdS black hole. The expansion of event horizon ceases, when the equipotential is approached. In the limit, the torsion becomes maximum and takes a critical value $P_{max}$. The maximal torsion $P_{max}$ is formally identified with the Nariai mass for dS. Thus the Hawking radiations, underlying the instabilities, transform a SdS to a Nariai black hole in the regime.
\begin{figure}
\centering
\mbox{\includegraphics[width=0.8\linewidth,height=0.3\textheight]{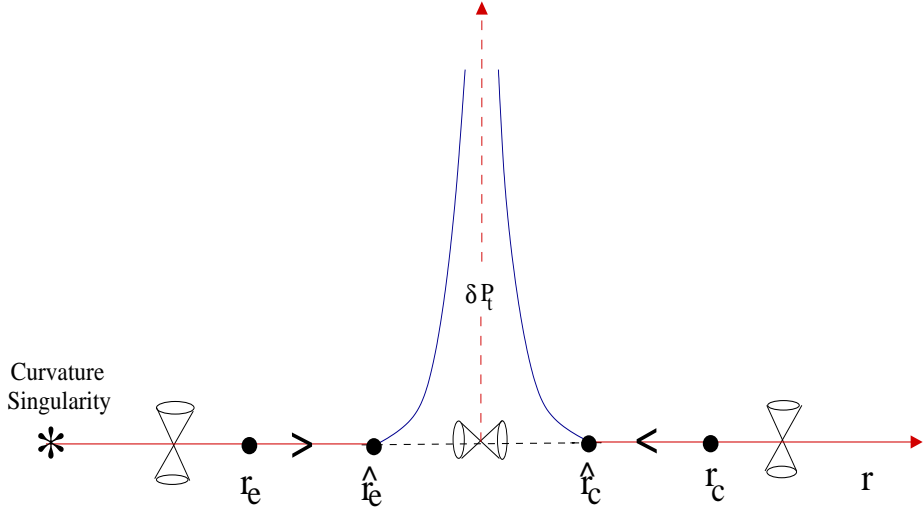}}
\caption{\it Schematic diagram shows a shock wave peak at ${\hat r}_e\leftrightarrow{\hat r}_c$, signifying a temporal phase $\delta P_t\neq 0$, in an emergent SdS black hole on an effective $D_4$-brane.}
\end{figure}

\sp
\noindent
In the context, an emergent TdS black hole (\ref{T-7}) may be argued to relate to a meta-stable phase of a near horizon Nariai black hole in the quantum geometry \cite{cai-myung-zhang,rong-cai}. The TdS black hole is described by a cosmological horizon at ${\tilde r}_{c}=(b+\delta P)$. In absence of a geometric torsion $P$, the topological geometry reduces to a pure dS with a cosmological horizon at $b$. However, the topological geometry ensures that an observer is at the other side of the horizon, which is in conformity with the cosmic censorship hypothesis. In fact, an emergent Nariai and an emergent TdS are different gauge choices due to different values of $C$ in the frame-work. Nevertheless, the independent geometries may be compared for their vacuum energies.

\sp
\noindent
Interestingly, the emergent vacua on an effective $D_3$-brane is essentially sourced by a geometric torsion ${\cal H}_3$ in a second order formalism. Alternately, the geometric torsion may also be viewed as a perturbation series (\ref{gauge-6}) in a gauge theory. At this point, we recall the significance of the coupling of a zero mode to non-zero mode in the gauge theory. It is important to note the role played by a non-zero mode in addition to a zero mode into the vacuum energy. A zero mode, in the open string metric (\ref{gauge-18}), is pumped into the vacuum to yield a non-zero potential energy, where a the quantum fluctuation in the vacuum are sourced by a non-zero mode. In other words, a fluctuation signifies a non-linear electro-magnetic field. Their local degrees do play a significant role to compute the total vacuum energy. It may qualitatively be analyzed from a generalized metric (\ref{gauge-32}), for its ${\tilde G}_{tt}$ component. The total energy function in various emerging vacua is worked out in a non-perturbative frame-work, underlying a geometric torsion, to yield 
\be
E(r)= \left ( {\tilde G}_{tt}-g_{tt} \right )\ .\qquad {}\label{T-8} 
\ee
Then, the energy function in an emergent SdS black hole, defined with a lorentzian signature, becomes
\be
E_S^l(r) = \left ( {{r^2}\over{b^2}} + {{P^6}\over{r^6}}\right )\ .\qquad {}\label{T-8a}
\ee
Similarly, the energy function in a TdS black hole with a lorentzian metric is given by
\be
E_T^l(r) = \left ( {{r^2}\over{b^2}} - {{P^6}\over{r^6}}\right )\ .\qquad {}\label{T-8b}
\ee
The energy underlying a Nariai black hole possesses a minimum at $r_0= (\sqrt{3}b)^{1/3}{\hat r}_e$. The vacuum energy at its spatial horizon ${\hat r}_c=b={\hat r}_e=P$ is estimated to yield
\bea
E_N^l(r)|_{{\hat r}_e}&=&\left ( 1 + {{P^2_{max}}\over{b^2}}\right )_P\nonumber\\
&\rightarrow& \left ( 1 + {{P^6_{max}}\over{b^6}}\right )_b\approx 1\ .\label{T-9}
\eea 
The vacuum energy in an emergent TdS, at cosmological horizon ${\tilde r}_c=(b+\delta P)$, becomes
\bea
E_T^l(r)|_{{\tilde r}_c}&=&\left ( 1 - {{P^6}\over{b^6}}\right )\ +\ {{2\delta P}\over{b}}\left (1 + {{3P^6}\over{b^6}}\right )\nonumber\\
&\rightarrow&\left ( 1 +\ {{2\delta P}\over{b}}\right ) \approx 1\ .\label{T-9a}
\eea 
It becomes evident that the energy in a Nariai vacuum is equal to that in a TdS vacuum and they are at the same spatial horizon $P_{max}(=b_{min}$). However, they are separated by a temporal phase $(\delta P)_t\neq 0$. With a subtlety, a Nariai vacuum may be identified with a TdS vacuum in quantum gravity. In other words, the maximal torsion in Nariai vacuum tunnels to a TdS vacuum in the quantum regime \cite{parikh,cai-myung-zhang,rong-cai,medved}. The maximal energy is used to change a negative gravitational mass in Nariai to a positive mass in TdS. The maximal torsion, may be identified with a condensate of discrete torsion in TdS vacuum. The SdS and TdS emergent black holes may arguably be believed to describe the near (cosmological) horizon $D_4$-brane geometries. Their cosmological horizon radii $(b-\delta P)$ and $(b+\delta P)$ further re-assures the near horizon brane geometry. A global view of the entire dS space-time in an effective curvature description may reveal a pluasible link in dS/CFT duality. An observer in SdS would likely to notice the Hawking radiations from both horizons in opposite directions to each other. A thermal analaysis in section 4.6 would ensure the direction of net flow towards its cosmological horizon. A positive energy particle, from a pair created just inside an event horizon and just out-side a cosmological horizon, travels to SdS. The gravitational energy in SdS increases via Hawking radiations.
As a result, a SdS evaporates and leaves behind a Nariai black hole. From the perspective of an observer in TdS, the negative energy (particle) from SdS tunnels to TdS through their near (cosmological) horizon boundaries. The positive energy anti-particle, from a created pair, inside the cosmological horizon raises the gravitational energy in SdS. The pair creation, just inside a cosmological horizon in TdS continues until the (dark) energy attains its maximum $P_{max}$ in SdS. In the limit, the cosmological scale reduces to its minimal value $b_{min}$, which is identifed with a condensate of discrete torsion.

\sp
\noindent
It is important to recall that an increase in torsion energy lowers the dS vacuum energy. A SdS black hole undergoes Hawking radiation to transform into a Nariai black hole defined with a maximum $P_{\max}$ in the emergent scenario. Thus the growth of dark energy ceases in a Nariai vacuum. The value $P_{max}$ is associated with a minimum in the cosmlogical scale $b$, which is indeed enforced by the phenomenon of Hawking radiation from a SdS black hole. Interestingly, a dynamical constant $P$, with its increasing value, gradually reduces the cosmological scale in the frame-work. A condensate of torsion is identified with a minimal cosmological scale $b_{min}$ (or with $P_{max}$) in TdS. The condensate in the topological phase gets cancelled, by the available cosmological vacuum energy, to lead to a typical $D_4$-brane underlying a gauge theory.

\begin{figure}
\centering
\mbox{\includegraphics[width=.95\linewidth,height=0.3\textheight]{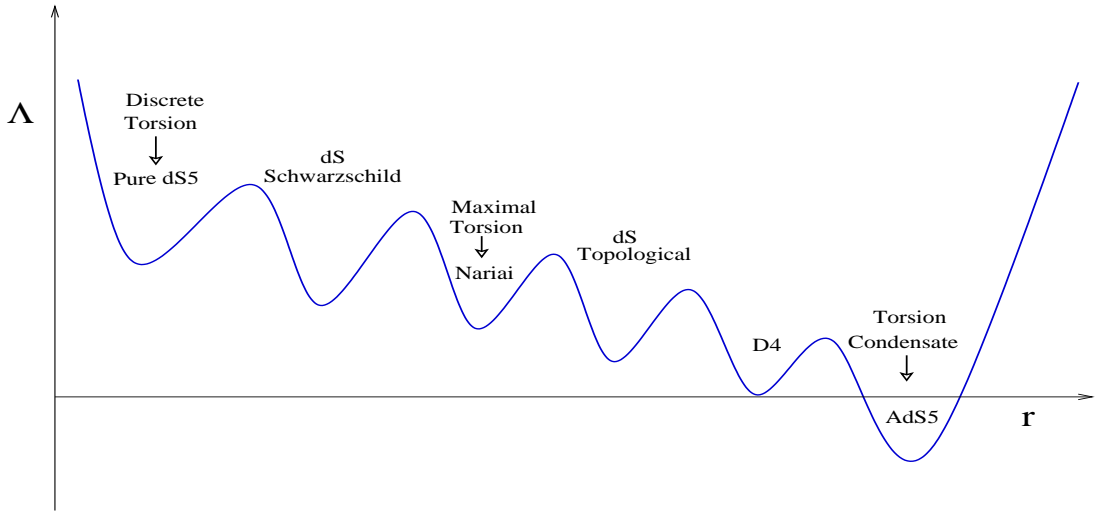}}
\caption{\it Schematic tunneling between the emerging vacua in a geometric torsion formulation on an effective $D_4$-brane. The dark energy, sourced by a discrete torsion, instigates Hawking radiations from a SdS black hole. A pure dS, in presence of the dark energy, tunnels all the way to an AdS via some of the meta-stable vacua. The maximal torsion forms a condensate in a TdS and tunnels to a stable AdS.}
\end{figure}
\subsection{AdS patch within thermal dS brane}
In this section, we primarily focus on an emergent metric signatures and its underlying energy function. We will observe that the dS vacua, defined by their energy functions, with a lorentzian signature turn out to be the correct description in the regime $P<r<b$ where as an euclidean AdS defines a correct vacuum for $b<r<P$ on an effective $D_4$-brane. An euclidean time in AdS enforces a field theoretic description to compute temperature for an AdS black hole. However, the thermal field theoretic tool may not be satisfactory to compute the temperature in lorentzian dS vacua. Nevertheless, the presence of a lower dimensional Schwarzschild AdS (SAdS) patch in a dS vacuum, underlying a brane world regime $P<r<b$ may be exploited to compute temperature in the dS geometries. The existence of an AdS in dS vacua may turns out to be insightful to the conjectured dS/CFT duality. 

\sp
\noindent
In the context, the tunneling geometries of meta-stable dS vacua in an effective $D_4$-bane formulation has been depicted in Figure 3. The tunneling has been argued to begin with the Hawking radiations in an emergent SdS. The radiations are further supplemented by a series of geometric transitions, presumably leading to a stable vacuum underlying an emergent AdS on an effective $D_4$-brane. Importantly, the role of a condensate $P_{max}$ in a TdS vacuum may be revisited in the regime $P<r<b$ to obtain an emergent AdS patch on an effective $D_4$-brane. 
\subsubsection{dS regime ($\mathbf{P<r<b}$):}
We begin with an emergent TdS black hole (\ref{T-7}) obtained with real time $t$. It may as well be expressed with an euclidean time $t_e=-it$. The relevant geometry may be re-expressed as:
\bea
&&ds^2= \left (1-{{r^2}\over{b^2}} + {{P^6}\over{r^6}}\right ) dt_e^2 + \left ( 1 -{{r^2}\over{b^2}} + {{P^6}\over{r^6}}\right )^{-1} dr^2 
+ 2i\left ( 1 + {{P^6}\over{r^6}}\right ) dt_e dr\qquad {}\nonumber\\
&&\qquad\qquad\qquad\qquad\qquad\qquad -\ {{2iP^6}\over{br^4}}\left ( dt_e -i dr \right )d\psi +\left ({{r^2}\over{b^2}}+ {{{P^6}}\over{b^2r^4}}\right ) 
r^2 d\Omega^2_3 \ .\label{qads-0}
\eea
Under an interchange $dt_e\leftrightarrow dr$, an euclidean geometry is analytically continued back to a lorentzian signature to yield
\bea
&&ds^2= -\left (1-{{r^2}\over{b^2}} + {{P^6}\over{r^6}}\right )^{-1} dt^2 + \left ( 1 -{{r^2}\over{b^2}} + {{P^6}\over{r^6}}\right ) dr^2 
+ 2\left ( 1 + {{P^6}\over{r^6}}\right ) dt dr\qquad {}\nonumber\\
&&\qquad\qquad\qquad\qquad\qquad\qquad -\ {{2iP^6}\over{br^4}}\left ( dr - dt \right )d\psi +\left ({{r^2}\over{b^2}}+ {{{P^6}}\over{b^2r^4}}\right ) 
r^2 d\Omega^2_3 \ .\label{qads-01}
\eea
In fact, the interchange of a spatial coordinate with a temporal is indeed dictated by the near horizon geometry on an effective $D_4$-brane in the frame-work. In the regime an effective $D_4$-brane, corresponding to the TdS balck hole, may appropriately be given by
\bea
ds^2= &-&\left (1+{{r^2}\over{b^2}} -{{P^6}\over{r^6}}\right ) dt^2 + \left ( 1 +{{r^2}\over{b^2}} - {{P^6}\over{r^6}}\right )^{-1} dr^2 
+ 2\left ( 1 + {{P^6}\over{r^6}}\right ) dt dr\nonumber\\
&-&\ {{2iP^6}\over{br^4}}
\left ( dr - dt \right )d\psi +\left ({{r^2}\over{b^2}}+ {{{P^6}}\over{b^2r^4}}\right ) 
r^2 d\Omega^2_3 \ .\label{qads-02}
\eea
At the first sight, the emergent TdS black hole has tunnelled into a SAdS black hole. In absence of a geometric torsion, an effective $D_4$-brane describes an $S^3$-symmetric AdS geometry. However, in presence of a torsion, the emergent SAdS possesses an imaginary angular momentum. Nevertheless, the emergent brane geometry becomes physical for a fixed $\psi$ $(={\pi}/2)$. Then, the reduced geometry becomes
\bea
ds^2= &-&\left (1+{{r^2}\over{b^2}} -{{P^6}\over{r^6}}\right ) dt^2 + \left ( 1 +{{r^2}\over{b^2}} - {{P^6}\over{r^6}}\right )^{-1} dr^2 
+ 2\left ( 1 + {{P^6}\over{r^6}}\right ) dt dr \nonumber\\
&+&\left ({{r^2}\over{b^2}}+ {{{P^6}}\over{b^2r^4}}\right ) 
r^2 d\Omega^2_2 \ .\label{qads-03}
\eea
A significant AdS patch of quantum geometry within a TdS vacuum is noteworthy. Under $r\rightarrow -r$, an effective ${\bar D}_3$-brane geometry may be obtained. Both, a brane and its anti-brane, describe an $S^2$-symmetric SAdS.

\sp
\noindent
Alternately, the unphysical angular momentum in the emergent black hole (\ref{qads-02}) may seen to decouple from the AdS brane in its asymptotic limit. As a result, an effective $D_4$-brane in its near horizon geometry is identified with an asymptotic AdS in the quantum regime. In an asymptotic limit, the emergent black holes on an effective $D_4$-brane and its anti-brane are given by
\be
ds^2= -{{r^2}\over{b^2}}dt^2\ +\ {{b^2}\over{r^2}} dr^2 \ \pm 2 dt dr\  +\ {{r^4}\over{b^2}} d\Omega^2_3 \ ,\label{qads-04}
\ee
where $b$ is identified with the AdS radius of curvature. With a global scenario, $i.e.$ a $D_4$-brane and a ${\bar D}_4$-brane together, a SAdS (\ref{qads-02}) within TdS is given by
\bea
&&ds^2= -\left (1+{{r^2}\over{b^2}} -{{P^6}\over{r^6}}\right ) dt^2 + \left ( 1 +{{r^2}\over{b^2}} - {{P^6}\over{r^6}}\right )^{-1} dr^2 \qquad\qquad\qquad {}\nonumber\\
&&\qquad\qquad\qquad\qquad\qquad\qquad +\ {{2iP^6}\over{br^4}} dt d\psi +\left ({{r^2}\over{b^2}}+ {{{P^6}}\over{b^2r^4}}\right ) 
r^2 d\Omega^2_3 \ .\label{qads-02a}
\eea
Under an analytic continuation $t=it_e$, followed respectively by an interchange $dr\leftrightarrow dt_e$ and an analytic continuation back to real time, the global SAdS becomes 
\bea
&&ds^2= -\left (1-{{r^2}\over{b^2}} +{{P^6}\over{r^6}}\right ) dt^2 + \left ( 1 -{{r^2}\over{b^2}} + {{P^6}\over{r^6}}\right )^{-1} dr^2 \qquad\qquad\qquad {}\nonumber\\ 
&&\qquad\qquad\qquad\qquad\qquad\qquad -\ {{2P^6}\over{br^4}} dr d\psi +\left ({{r^2}\over{b^2}}+ {{{P^6}}\over{b^2r^4}}\right ) 
r^2 d\Omega^2_3 \ .\label{qads-02b}
\eea
Interestingly, the global geometry with a charge corresponds to that in a global TdS obtained from eq.(\ref{T-7}) with an angular momentum. It is evident that the magnitude of a charge in a TdS is equal to the magnitude of an angular momentum in SAdS. Thus, from a global perspective, 
a TdS vacuum  may describes a SAdS patch and vice-versa. The aspects of a Nariai black hole, on an effective $D_4$-brane, may be recalled to identify $r\rightarrow b=P_{max}$, spatially, at the Nariai horizon. In other words, a condensate of torsion may also be identified with an AdS radius. The emerging phenomenon of an AdS patch within dS, in the frame-work, needs further attention. 

\sp
\noindent
On the other hand, the energy function for an emerging AdS, within a dS, may be worked out with an euclidean and a lorentzian metric signatures to yield
\be
{\hat E}^e_{AdS}(r)=-{\hat E}^l_{AdS}(r)= \left ( {{r^2}\over{b^2}}-{{P^6}\over{r^6}}\right )\ .\qquad\qquad {}\label{03a}
\ee
Interestingly, the energy function for an euclidean AdS identifies with a lorentzian TdS (\ref{T-8b}) in the regime. It re-assures the presence of an AdS geometry within a dS vacuum on an effective $D_4$-brane. In addition, the formal identification possibly signals a signature
change in the effective metric geometries. The correct vacuum for an euclidean and a lorentzian AdS brane may seen to be defined, respectively, at $P_{max}$ and at $b_{min}$. Explicitly the AdS vacuum energies, within a dS topological, are given by
\bea
&&{\hat E}^e_{AdS}(r)|_P=\left ( {{P^2}\over{b^2}} - 1\right )<0\ ,\qquad\qquad {}\nonumber\\
&&{\hat E}^l_{AdS}(r)|_b=\left ( {{P^6}\over{b^6}} - 1\right )<0\ .\label{03b}
\eea
The fact that $P_{max}=b_{\min}$, for a spatial radial coordinate, at the horizon in a Nariai phase eliminates the apparent difference in the horizon radius in the AdS patch. In other words the difference between the temporal radii, at the event and cosmological horizons in a Nariai vacuum, enforces
a signature change in an AdS metric geometries within a dS topological vacuum. The tunneling of Nariai quantum geometry to a dS topological phase essentially imply that an AdS causal patch is indeed a significant phase in SdS or generically in a dS vacuum in the frame-work.

\subsubsection{AdS regime ($\mathbf{b<r<P}$):}
At this point, it may be insightful to perform an alnalysis for an allowed range $b<r<P$ with $b\neq 0$ and $P\rightarrow P_{max}$. The quantum geometry on an effective $D_4$-brane is worked out ab initio using another Weyl scaling of the emergent metric (\ref{gauge-22}). It is given by
\be
{\tilde G}_{\mu\nu}= {{P^6}\over{(2\pi\alpha')^2r^2}} {{\tilde G}''}_{\mu\nu}\ .\qquad\qquad {}\label{qads-1}
\ee
The large conformal factor, in the allowed range, ensures a quantum geometry underlying an emergent metric ${{\tilde G}''}_{\mu\nu}$ on an effective $D_4$-brane. It further re-assures the small $r$ limit in the freame-work. The transformed metric with $C=\pm (1/2)$ in an effective theory, may lead to some interesting geometries in a second order formalism when $r^4$$>$$(bP^3)$. In the limit, an effective $D_4$-brane line-elements take elegant forms and are given by
\bea
&&ds^2=-\left (- l^2 + {{r^2}\over{p^2}} + {{Q^6}\over{r^6}} \right ) dt^2\ +\ 
\left (-l^2 +{{r^2}\over{p^2}} + {{Q^6}\over{r^6}}\right )^{-1} dr^2\ +\ 2\left (l^2 - {{Q^6}\over{r^6}} \right ) dt dr\qquad {}\nonumber\\
&&\qquad\qquad\qquad\qquad\qquad\qquad + {{(2\pi\alpha')Q^3}\over{r^4}}\left ( dt + dr \right )d\psi +\left ( {{r^2}\over{p^2}} - {{(2\pi\alpha')^2}\over{r^4}}\right ) r^2 d\Omega^2_3\label{qads-2}
\eea
\vspace{-.1in}
and
\vspace{-.1in}
\bea
&&ds^2=-\left (- l^2 + {{r^2}\over{p^2}} - {{Q^6}\over{r^6}} \right ) dt^2\ +\ 
\left (-l^2 +{{r^2}\over{p^2}} - {{Q^6}\over{r^6}}\right )^{-1} dr^2\ +\ 2\left (l^2 + {{Q^6}\over{r^6}} \right ) dt dr\qquad {}\nonumber\\  &&\qquad\qquad\qquad\qquad\qquad\qquad -{{(2\pi\alpha')Q^3}\over{r^4}}\left ( dt + dr \right )d\psi +\left ( {{r^2}\over{p^2}} + {{(2\pi\alpha')^2}\over{r^4}}\right ) r^2 d\Omega^2_3\ ,\label{qads-3}
\eea
where $l^2= b^2/p^2<<1$, $p=P^3/(2\pi\alpha')$ and $Q^3=(2\pi\alpha' b)$ are non-zero constants. The regime may be viewed under an interchange of the dynamical constants ($P\leftrightarrow b$) when compared with the dS regime on an effective $D_4$-brane. Here $P$ and $b$, respectively, correspond to
a cosmological scale and an energy scale. The vacuum energy density (\ref{ads-9}) confirms ${\tilde\Lambda}<0$, and becomes extremely small, in the regime for both the vacua. The emergent brane geometries may be identified with two distinct AdS (AdS${}^{\pm}$) geometries in small $r$. Interestingly, the limit $r^4$$>$$(bP^3)$ may numerically be investigated to define, a priori, two geometric ranges: $0<b<1$ or small $b$ and $b\ge 1$ or large $b$ within $b<r<P$ on a brane. A stable AdS${}^-$ possesses an event horizon at $r_h=(b+\delta b)$ for small $b$. The event horizon in AdS${}^-$ moves to $r_h=(p-\delta b)$ for large $b$ in the regime. Nevertheless, the AdS${}^-$ geometry on an effective $D_4$-brane may describe a SAdS black hole. AdS${}^+$ does not possess a real horizon in the coordinate system. The computation of angular velocity for an emergent SAdS (\ref{qads-3}) at the event horizon for small $b$ yields $\Omega_h= (1/b)$ becomes large.  For large $b$, the angular velocity becomes infinitely large at the horizon. Thus the angular velocity does not distinguish between a small $b$ and large $b$ geometric ranges in the frame-work. The emergent SAdS black hole (\ref{qads-3}) simplifies in the asymptotic limit to yield
\be 
ds^2=- {{r^2}\over{p^2}} dt^2\ +\ {{p^2}\over{r^2}} dr^2\ \pm\ 2 l^2\ dt dr\
+\ {{r^4}\over{p^2}}\  d\Omega^2_3\ .\label{qads-4}
\ee 
In the limit, the effective $D_4$-brane geometry retains the spherical symmetry. The redefined charge $p$, signifying the presence of a dynamical torsion, may be identified with an AdS radius of curvature in a strongly coupled regime. Eqs.(\ref{qads-04}) and (\ref{qads-4}), respectively, defined by $P<r<b$ and $b<r<P$ ensure a subtle flip between $b\leftrightarrow P$ in the frame-work. Thus, an effective $D_4$-brane in the quantum regime may as well be described by an SAdS microscopic black hole. The geometric tunneling essentially interchanges a condensate, sourced by a local mode, with a global mode in a discrete torsion. Unlike to the quantum dS black hole (\ref{dS-2}), the emergent SAdS black hole is primarily governed by a global mode of two form which in turn yields a topological torsion in the theory. The absence of local torsion in the quantum geometry re-assures a stable vacuum in the regime. 

\sp
\noindent
In absence of a torsion, the corresponding black hole underlying a string theory may seen to be defined with a vanishing cosmological constant and also with a vanishing source energy ($0<r<\infty$). A priori, the vacuum geometry is given by
\be
ds^2=  l^2 dt^2\ -{1\over{l^2}}dr^2\ +\ 2l^2\ dt dr\ +\ {{(2\pi\alpha')^2}\over{r^2}} d\Omega^2_3\ .\label{qads-6}
\ee
The $(-)$ve sign in the metric determinant signifies lorentzian signature in the near horizon brane geometry. Under $dr\leftrightarrow dt$ in the near horizon, the $D_4$-brane explicitly re-assures a real time. After the interchange, the near horizon line element may be given by
\be
ds^2= -{1\over{l^2}} dt^2\ +\ l^2 dr^2\ +\ 2l^2\ dt dr\ +\ {{(2\pi\alpha')^2}\over{r^2}} d\Omega^2_3\ .\label{qads-7}
\ee
\begin{figure}
\centering
\mbox{\includegraphics[width=0.7\linewidth,height=0.22\textheight]{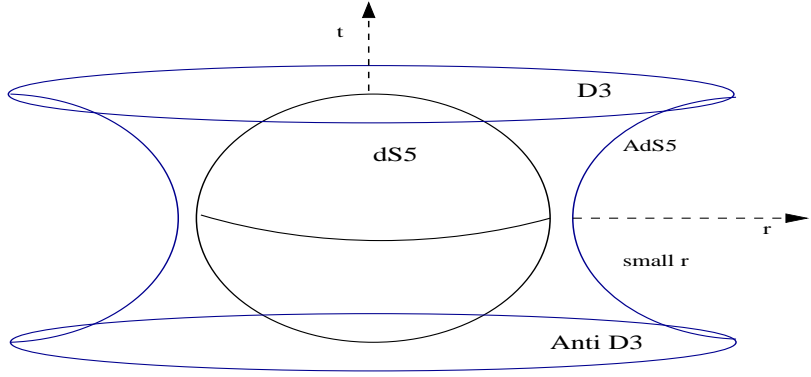}}
\caption{\it The emergent dS geometry on an effective $D_4$-brane, in small $r$, may be viewed as a closed Universe. Under a series of geometric transitions, leading to a stable vacuum, the primordial de Sitter geometry in its topological phase convolutes to an asymptotic AdS. The effective $D_4$-brane, underlying an emerging AdS, may be viewed as a pair of $D_3$-brane and anti-brane in the quantum regime.}
\end{figure}

\vspace{-.1in}
\noindent
The emergent SAdS energy function (\ref{T-8}) in the regime $b<r<P$ is computed for the near horizon ($r_h\pm \epsilon$) brane. The energy function for the near horizon ($r_h+\epsilon$) may a priori be given by
\be
E_{AdS}^{e+}(r)=\left ({{r^2}\over{P^6}} - {{b^2}\over{r^6}}\right )\ ,\qquad {}\label{qads-8}
\ee
where we set $(2\pi\alpha')=1$ for simplicity. The energy function is defined prior to a flip in light cone at the horizon. In the limit $r\rightarrow r_h=b$, an AdS vacuum energy becomes
\be
E_{AdS}^{e+}(r)|_b=-\left ( {1\over{b^4}} - {{b^2}\over{P^6}}\right )<0\ .\qquad {}\label{qads-9}
\ee
The energy function after a flip in light cone in the near horizon ($r_h-\epsilon$) regime may a priori 
be re-expressed as
\be
E_{AdS}^{e-}(r)=\left ( -{{b^2}\over{P^6}} - {{P^6}\over{b^2}} + {{r^2}\over{P^6}} - {{b^2}\over{r^6}}\right )\ .\label{qads-10}
\ee
In the limit $r\rightarrow r_h=b$, the vacuum energy in the near horizon ($r_h-\epsilon$) becomes
\be
E_{AdS}^{e-}(r)|_b= -\left ({{P^6}\over{b^2}} +{1\over{b^4}}\right )<0\ .\qquad {}\label{qads-11}
\ee
The extremely small value of vacuum energy of an emergent black hole re-assures an euclidean SAdS on an effective $D_4$-brane in its near horizon ($r_h\pm\epsilon$). The euclidean signature of the emergent SAdS may also be reconfirmed directly at its horizon $r_h=b$ by computing the AdS energy function (\ref{T-8}) with a flat metric $g_{\mu\nu}$. The energy function is given by
\be
E_{AdS}^{e}(r)=\left ( -1 -{{b^2}\over{P^6}} + {{r^2}\over{P^6}} - {{b^2}\over{r^6}}\right )\ .\label{qads-12}
\ee
At the horizon, the vacuum energy becomes
\be
E_{AdS}^{e}(r)|_b= -\left (1 +{1\over{b^4}}\right )<0\ .\qquad {}\label{qads-13}
\ee
The SAdS vacuum energy in a limit $r\rightarrow r_h$ may be arranged as:
\be
E_{AdS}^{e+}|_b>E_{AdS}^{e}|_b>E_{AdS}^{e-}|_b\ .\label{qads-14}
\ee
Thus for small $b$ the continuity in vacuum energy in a near horizon black hole presumably enforces an euclidean SAdS geometry. This in turn allows one to compute the AdS temperature using a field theoretic prescription in the regime for all values of $b$.

\subsection{Thermal equilibrium}
The Hawking temperature in the emergent SAdS black hole (\ref{qads-3}), at its event horizon $r_h$, is computed in the AdS regime to yield
\be
T^{\rm AdS}={1\over{2\pi r_h}}\left ( {{4r_h^2}\over{p^2}} - 3l^2\right )\ .\qquad {}\label{qads-15} 
\ee
Using a plausible unit $T_0=1/(2\pi p_{max})$, the temperature of SAdS on an effective $D_4$-brane becomes 
\be
T^{\rm AdS}=nT_0\left ( {{{4r_h^2} - {3b^2}}\over{pr_h}} \right )\ .\qquad {}\label{qads-16}
\ee
For small energy scale $0<b<1$, the event horizon in an emergent SAdS turns out to be at ($b+\delta b$). The temperature may easily be compared by the relative scale factors in the quantum regime. A priori, the temperature in an emergent SAdS black hole (\ref{qads-03}) on an effective $D_4$-brane may explicitly be given by
\bea
T_{b<1}^{\rm AdS}&=&{1\over{2\pi b}}\left ( 1 + {{7\delta b}\over{b}}\right ){{b^2}\over{p^2}}\qquad {}\nonumber\\
&=&nT_0\left ( 1 + {{7\delta b}\over{b}}\right ){{b}\over{p}}\nonumber\\
&\approx& nT_0 \left ({{b}\over{p}}\right )\ .\label{qads-14a}
\eea
With large $b\ge 1$, the SAdS black hole event horizon moves to ($p-\delta b$). The Hawking temperature in the regime becomes
\bea
T_{b\ge 1}^{\rm AdS}&=&{1\over{2\pi p}}\left ( 4 \left ( 1 -{{\delta b}\over{p}}\right ) - 
3\left (1 + {{\delta b}\over{p}}\right ){{b^2}\over{p^2}}\right )\qquad\qquad {}\nonumber\\
&\rightarrow& 4nT_0\left ( 1 -{{\delta b}\over{p}}\right )\nonumber\\
&\approx& mT_0\ .\label{qads-16a}
\eea
The drop in scale factor in the temperature with large $b$, when compared with that in small $b$, a priori implies $T_{b\ge 1}^{\rm AdS}>T_{b<0}^{\rm AdS}$ on an effective $D_4$-brane. However, incorporating an underlying tunneling between the dS vacua and an AdS vacuum, the ratio of scale factor $b/p\rightarrow 1$ in the Nariai phase. Thus, the temperature of a SAdS black hole, for all $b$, becomes $T_0$.

\sp
\noindent
Interestingly the AdS temperature computed in a field theoretic technique, may be used to obtain the temperature ${\tilde T}^{\rm AdS}$ in an AdS patch within a dS vacua in the regime $P<r<b$. Nevertheless, we perform thermal analysis for an emergent SAdS black hole (\ref{qads-3}), within dS to compute the temperature. It is given by
\bea
{\tilde T}^{\rm AdS}&=&{1\over{2\pi r_h}}\left ( {{4r_h^2}\over{b^2}} + 3\right )\qquad {}\nonumber\\
&=&nT_0\left ({{p}\over{b}}\right ) \left ( {{{4r_h^2} + 3b^2}\over{br_h}} \right )\ .\qquad {}\label{qads-17}
\eea
Thus the temperature of an emergent SAdS black hole at its event horizon $(p+\delta p)$ is 
\bea
{\tilde T}^{\rm AdS}&=&nT_0\left ( 3 \left ( 1 -{{\delta p}\over{p}}\right ) + 4\left ( 1 + {{\delta p}\over{p}} \right ) {{p^2}\over{b^2}}\right ) 
\nonumber\\ &\rightarrow &3n T_0\left (1 - {{\delta p}\over{p}}\right )\nonumber\\
&\approx& mT_0\ .\label{qads-18}
\eea
Eqs.(\ref{qads-16a}) and (\ref{qads-18}) confirm that the temperature in an emergent SAdS black hole is equal to that of an SAdS within a TdS. 
It implies a thermal equilibrium between an AdS vacuum and a TdS vacuum in the quantum regime. However, the temperature in TdS is computed at its cosmological horizon ($b-\delta p$), where as the temperature (\ref{qads-18}) in an AdS is obtained at an event horizon. Thus a field theoretic technique may as well be applied to compute the temperature in an emergent TdS (\ref{T-7}) at its cosmological horizon to yield
\bea
{\tilde T}^{\rm TdS}&=&{1\over{4\pi}}|\partial_r{{\tilde G}''}_{tt}(r){\Big |}_{r_c}\nonumber\\
&=&{1\over{2\pi r_c}}\left ({{4r_c^2}\over{b^2}} -3 \right )\ ,\label{qads-19}
\eea
where the mod ensures a positive temperature at $r_c$. However, the temperature expression in TdS appears to be different than AdS (\ref{qads-17}) in the dS regime. The temperature of an emergent TdS black hole may explicitly work out to yield
\bea
{\tilde T}^{\rm TdS}&=&n{T_0}\left ( 1-{{7\delta p}\over{b}}\right )\left ({p\over{b}}\right )\nonumber\\
&\approx& n T_0\left ({p\over{b}}\right )\ .\label{qads-20}
\eea
A priori, the temperature ${\tilde T}^{\rm TdS}<{\tilde T}^{\rm AdS}$. However, an identification of an emergent TdS black hole 
with  Nariai black hole uses  Nariai limit $p\rightarrow b$, In the limit,
the SAdS temperature (\ref{qads-18}) turns out to be equal to the computed TdS temperature (\ref{qads-20}). The temperature relation becomes ${\tilde T}^{\rm AdS}=T_0={\tilde T}^{\rm TdS}$. Furthermore, a thermal equilibrium between an emergent SAdS and a TdS black hole(s) may allow one to use the thermal field theoretic techqnique to compute the temperature in the emergent SdS black hole at its event horizon. The temperature in SdS at its event horizon ($p+\delta p$) is given by
\bea
{\tilde T}^{\rm SdS}_{r_e}&=&{1\over{4\pi}} |\partial_r{{\tilde G}''}_{tt}(r){\Big |}_{r_e}\nonumber\\ 
&=&{1\over{2\pi r_e}}\left ( 3-{{4r_e^2}\over{b^2}} \right )\nonumber\\
&=&nT_0\left ( 3\left (1 -{{\delta p}\over{p}}\right ) - 4\left ( 1 + {{\delta p}\over{p}} \right ) {{p^2}\over{b^2}}\right )\nonumber\\
&\rightarrow&3nT_0\left ( 1 -{{\delta p}\over{p}}\right )\nonumber\\
&\approx& mT_0\ .\label{qads-21}
\eea
The temporal phase $\delta P_t\neq 0$ in SdS does not seem to validate the thermal field theoretic technique to compute temperature in a SdS at its cosmological horizon $r_c$. Nevertheless, under a change in signature the SdS vacuum may be identified with a TdS vacuum. Thus the temperature in SdS at $r_c$ may be obtained using eq.(\ref{qads-19}). However, the notional change in metric signature may provoke thought to re-express the temperature at $r_c$ in a generic dS using a lorentzian metric without a mod in its formal definition. 
Generically, a mod ensures a positive temperature at the horizon. It may be relaxed at the expense of a lorentzian metric at $r_c$. The notion may further be supported by an observation that the geometries were euclidean or hot at $r_c$. Presumably, the notion of time becomes significant with a $D_4$-brane and ${\bar D}$-brane under a generalized description of branes within a brane. Intuitively, the temperature in an SdS at its cosmological horizon may, equivalently, be given by
\bea
{\tilde T}^{\rm SdS}_{r_c}&=&-{1\over{4\pi}} \partial_r{{\tilde G}''}_{tt}(r){\Big |}_{r_c}\nonumber\\
&=&{1\over{2\pi r_c}}\left ({{4r_c^2}\over{b^2}} -3\right )\ .\label{qads-22}
\eea
The temperature expression holds good for its evaluation at the cosmological horizon. The non-zero temporal phase in pure dS and SdS are primarily responsible for the renewed definition. With a subtlety, it may imply that the dS vacua are generically associated with a high temperature presumably with a ``real time'' and may enhance our understanding of dual euclidean CFT. We compute the temperature in the emergent SdS,  at its cosmological horizon ($b-\delta p$). It becomes
\bea
{\tilde T}^{\rm SdS}_{r_c}&=&nT_0\left ( 1 -{{7\delta p}\over{b}}\right )\left ({p\over{b}}\right )\nonumber\\
&\approx& mT_0\left ({p\over{b}}\right )\ .\label{qads-23}
\eea
It is evident that the temperature in an emergent SdS, at $r_c$, is supressed by a scale factor $(p/b)<1$, than its temperature estimated at $r_e$. It re-assures a net flow, of Hawking radiations, towards $r_c$ in SdS. As a result, the vacuum would eventually evolve towards a pure dS. The phenomenon is in agreement with the second law of thermodynamics underlying the fact that the total entropy in an asymptotic SdS is bounded from above by the entropy of a pure dS.

\sp
\noindent
In the Nariai limit, the scale factor becomes unity to re-assure a thermal equilibrium between the two horizons in SdS.
The Hawking temperature at $r_c$ in SdS (\ref{qads-22}) precisely identifies with the temperature obtained (\ref{qads-15}) for SAdS in Nariai limit. Thus the emergent SdS, TdS, AdS within TdS and AdS black holes may seen to be in thermal equilibrium at a minimal temperature $T_0$ in Nariai limit. Interestingly, the scale factor plays a significant role in the emergent dS and AdS black holes on an effective $D_4$-brane. 
A small scale factor associated with the computed temperature in dS and AdS regime(s), presumably imply that the stringy vacua are hotter than the brane-world or emergent vacua. An increase in (dark) energy lowers the temperature in dS vacua. The temperature attains its minimal value $T_0$, where various phases of the emergent dS vacua are in thermal equilibrium with a stable AdS vacuum. The role of $P$ and $b$ are, respectively, in striking analogy with a non-linear electric field and a magnetic field on a $D$-brane, in presence of a zero mode.

\sp
\noindent
On the other hand, the energy function for an emergent AdS with lorentzian metric is worked out in the near horizon ($r_h\pm\epsilon$) brane. A lorentzian SAdS (\ref{qads-3}) in the regime is described for $b\ge 1$ in the limit $r\rightarrow (r_h+\epsilon)=(P+\epsilon)$. However, after a flip of light cone in the near horizon, $i.e.$ in the limit $r\rightarrow (r_h-\epsilon)=(P-\epsilon)$, a SAdS  is surprisingly described by an euclidean metric signature. The change in metric signature \cite{kar-majumdar3} in the near horizon AdS brane may be a consequence of electric-magnetic duality for the source field in the gauge theory. The change in metric signature in the near horizon SAdS black hole needs further attention and is beyond the scope of this paper. Nevertheless, a naive computation of vacuum energy for a lorentzian SAdS at its near horizon $(P-\epsilon)$ may unfold the mystery behind a signature change. The vacuum energy confirms a geometric transition AdS $\leftrightarrow dS$ on an effective $D_4$-brane. The transition in the near horizon brane geometries may be viewed at the expense of a change in metric signature there.

\sp
\noindent
At this point, let us recall that a SAdS ($b<r<P$) and a TdS ($P<r<b$) possess their horizon at $r_h=b$, though a SAdS is defined at $r_e$
and a TdS is defined at $r_c$. It is important to note the same horizon radius $b$ for two distinct quantum geometries. Furthermore an AdS patch, within a TdS, is defined with its event horizon at $b$. Intuitively, the dynamical constant $b$ plays a significant role in the quantum geometries. 

\sp
\noindent
In the context, the obtained AdS vacuum energies (\ref{qads-11})-(\ref{qads-13}) are in agreement with the theory (\ref{ads-9}). The vacuum energy in dS vacua (\ref{T-9}) and (\ref{T-9a}) may be compared with the AdS to re-assure a tunneling of a TdS to a SAdS via a $D_4$-brane. The quantum dS vacuum (\ref{T-7}) presumably undergoes a convolution to unfold a hyperbolic geometry to an AdS vacuum (\ref{qads-3}). The nontrivial contribution from a local torsion, in the disguise of dark energy, increases at a faster rate in a dS  geometry. The growth of energy incorporates instabilities into the quantum dS. As a result, the discrete torsion tunnels to an AdS vacuum within the purview of a quantum theory. The stability of SAdS is re-assured by a condensation of torsion quanta to its maximum (critical) value. The maximum value of $P$ naturally dictates a smallest value for $b$ in the small $r$ regime. Interestingly the role of a discrete torsion, in an emergent dS geometry, gets interchanged with a global mode of a two form in a quantum AdS. Under a strong-weak coupling duality in the theory, the quantum AdS brane further maps to a macroscopic extremal AdS geometry (\ref{ads-8a}) obtained in large $r$. 
\section{Concluding remarks}
To summarize, we have investigated certain features of a non-zero mode of NS-NS two form in an open string world-sheet dynamics. Firstly, we have
explored an effective curvature dynamics, underlying a propagating torsion, on a $D$-brane gauge theory. Secondly, we have analyzed the emergent vacua on an effective $D_4$-brane for its tunneling. In the context, we have attempted to address the beginning of a $D_4$-brane Universe with a Big Bang, which is followed by a creation of a $D$-instanton.

\sp
\noindent
In the pretext of an effective curvature, we have revisited the $U(1)$ gauge theory on a $D_4$-brane in presence of a non-zero mode in the string bulk. The covariant derivative was modified iteratively by the gauge connections to yield an exact appropriate derivative in the gauge theory. Interestingly, the covariant derivative, exact in two-form, in a perturbation theory acts as a non-perturbative covariant derivative in a geometric formulation underlying a second order formalism. The $U(1)$ gauge invariance of the modified field strength, under a two form transformation, is exploited to obtain an intrinsic metric fluctuation governed by a geometric torsion. In fact, a generalized effective metric is sourced by a non-zero mode in the string world-sheet dynamics. The propagation of a geometric torsion, underlying an effective $D_4$-brane, was appropriately addressed in a second order formalism which in turn was shown to be described by a fourth order curvature tensor ${\cal K}_{\mu\nu\lambda\rho}$. Importantly, the generalized curvature reduces to the Riemann curvature tensor for a topological torsion in the theory. In the context, we have briefly out-lined the significance of a BTZ black hole \cite{btz} on an effective $D_2$-brane underlying the generalized curvature ${\cal K}_{\mu\nu\lambda\rho}$ in the formalism. However, it remains to check the construction of an emergent three dimensional dS geometry\cite{banados} along with a BTZ black hole underlying an effective $D_2$-brane.

\sp
\noindent
On the other hand, a two form ansatz in the gauge theory is used to construct an appropriate geometric torsion in the second order formalism underlying a generalized curvature scalar ${\cal K}$. An effective $D_4$-brane, underlying a near horizon geometry of a black hole, is shown to describe an asymptotic AdS in the semi-classical regime. Interestingly, a non-zero mode becomes insignificant in the large $r$ regime. In other words, the local modes of torsion decouple from the effective geometry and hence an effective $D_4$-brane is described by a topological torsion. A zero-mode 
corresponds to an AdS radius of curvature and hence an effective $D_4$-brane may be viewed as an AdS-brane. The existence of an effective anti $D_4$-brane in the frame-work has been argued, under $r\rightarrow -r$. However, a $D_4$-brane is at an inaccessible spatial distance away from a ${\bar D}_4$-brane. Nevertheless, the angular momentum in an effective $D_4$-brane is nullified by that in its anti-brane. In presence of an hypothetical extra transverse (spatial) dimensions, together they may describe an effective $D_5$-brane.

\sp
\noindent
We have investigated the small $r$ regime on an effective $D_4$-brane. Emergent geometries have been shown to describe dS black holes for $P<r<b$ and an SAdS black hole for $b<r<P$. Surprisingly, the dS-vacua were found to be mixed with Schwarschild and topological causal patches. They were separated out under a generalized matrix, for a discrete transformation. As a result, we have obtained a SdS and a TdS black holes in a quantum regime. The potential difference, at $r_e$ and $r_c$, set up instability in a SdS black hole. As a result, the emergent SdS black hole undergoes Hawking radiation. Its mass (or energy), underlying a discrete torsion, increases and finally the energy approaches a maximum $P_{max}$ value and the reduced dS geometry has been identified with a Nariai black hole in the quantum regime. The growth of $P$ ceases in Nariai vacuum and the $P_{\max}$ forms a condensate of discrete torsion. The energy function analysis identifies the energy condensate in Nariai vacuum with a cosmological horizon radius in a TdS. Interstingly, the condensate tunnels down to an euclidean AdS vacuum via TdS  and a typical $D_4$-brane, when $P<r<b$. Our analysis revealing an existence of SAdS patch within a TdS in the quantum regime is new and needs further attention. On the other hand the vacuum energy computed for an SAdS black hole with lorentzian signature in a different regime ($b<r<p$), was analyzed in the near horizon brane geometry to reveal a plausible metric signature change in the frame-work.
It was shown that a geometric transition SdS $\leftrightarrow$ TdS, in a near horizon $D_4$-brane, may alternately be viewed at the expense of a change in signature. The vacuum energy in the regime ($b<r<P$) was exploited to show the existence of an euclidean SAdS vacuum on an effective $D_4$-brane. Hawking temperature using a thermal field theoretic prescription in the regime was computed for a SAdS at its horizon(s). The temperature in AdS ($b<r<P$) is analyzed together with that in an emergent AdS within a dS regime ($P<r<b$), to obtain a formal expression for temperature in dS. It was shown that a SdS vacuum is defined with a higher temperature than than the AdS vacuum. Hawking radiations establishes a thermal equilibrium between a Nariai, TdS and SAdS in the formulation. Further investigation underlying the tunneling geometries between a hot dS and an AdS may reveal clue to unfold dS/CFT correspondences.

\sp
\noindent
Interestingly, the high temperature in an emergent dS has been the key to the origin of an effective $D_4$-brane (${\bar D}_4$-brane) Universe with a Big Bang (Crunch). It was argued that a pair $(D{\bar D})_{-1}$ was instantaneously created, by a discrete (light) torsion, at the Big Bang singularity. They moved away from each other along an inaccessible radial ($+r$ and $-r$) coordinates, which in turn gave birth to a pair of particle $(D{\bar D})_0$. The world-line for a $D_0$-brane was interpreted as the radius of an $S^2$. An emerging $D$-string described by the $S^2$, together with a $D_0$-particle, were argued to shape an effective $D_2$-brane at the cosmological horizon. Similarly, their anti-branes lead to an effective ${\bar D}_2$-brane in the frame-work. A significant geometry was argued to be created with a $D_2$-brane (${\bar D}_2$-brane), where the torsion is topological. A $D_2$-brane along with its anti-brane, in the strong coupling, qualitatively describes an effective $D_3$-brane (${\bar D}_3$-brane). Subsequently, a $(D{\bar D})_3$ pair nucleates an effective $D_4$-brane in the regime. The analysis may hint to a generalized notion of branes within a brane in the frame-work. In fact, the near horizon effective $D_p$-brane geometry collectively generalizes the branes within branes established independently for even and odd $p$-branes. A non-zero mode has been vital to a generalized description of all lower branes within an effective $D_4$-brane, in its near horizon geometry.

\sp
\noindent
In the context, we have investigated some aspects of an effective $D_3$-brane obtained from an effective curvature theory on a $D_4$-brane. A non-linear magnetic charge may be generated without the notion of a magnetic point charge defined by a geometric torsion. In fact, the non-linear magnetic charge may seen to be constructed from an electric point charge in presence of a local torsion. The detail is beyond the scope of this paper and is in progress by the authors of this paper.

\section*{Acknowledgments}
We gratefully thank K.S. Narain, Sudhakar Panda, S. Randjbar-Daemi, Swarnendu Sarkar, John H. Schwarz and Ashoke Sen for useful discussions at various stages of the work. A preliminary version of the work was presented by A.K.S. in an International Conference on Light Cone Physics Delhi 2012 December 10-15. A.K.S. acknowledges CSIR and S.S. acknowledges UGC for their fellowship. The work of S.K. is partly supported by a research grant-in-aid under the Department of Science and Technology, Govt.of India.

\def\anp{Ann. of Phys.}
\def\prl{Phys.Rev.Lett.}
\def\prd#1{{Phys.Rev.}{\bf D#1}}
\def\jhep{JHEP\ {}}{}
\def\cqg#1{{Class.\& Quant.Grav.}}
\def\plb#1{{Phys. Lett.} {\bf B#1}}
\def\npb#1{{Nucl. Phys.} {\bf B#1}}
\def\mpl#1{{Mod. Phys. Lett} {\bf A#1}}
\def\ijmpa#1{{Int.J.Mod.Phys.}{\bf A#1}}
\def\mpla#1{{Mod.Phys.Lett.}{\bf A#1}}
\def\rmp#1{{Rev. Mod. Phys.} {\bf 68#1}}


\end{document}